\documentclass[sigconf,nonacm,screen]{acmart}
\usepackage{array}
\usepackage{placeins}
\usepackage{etoolbox}
\graphicspath{{figures/}{./}}
\setcopyright{none}
\acmDOI{}
\acmISBN{}
\date{}
\hypersetup{pdfauthor={Yuxuan Weng and Yunge Wen},pdfsubject={Research preprint},hidelinks}
\setcitestyle{numbers,sort&compress}

\AtBeginDocument{\raggedbottom}
\begin{document}

\title{NephoCodex: Exploring Bounded Material Agency in Weather Data Physicalization}
% Author details supplied by Yuxuan Weng on 2026-09-15.
\author{Yuxuan Weng}
\authornote{Corresponding author.}
\affiliation{\institution{Nanyang Technological University}\city{Singapore}\country{Singapore}}
\email{WENG0051@e.ntu.edu.sg}
\author{Yunge Wen}
\affiliation{\institution{Massachusetts Institute of Technology}\city{Cambridge}\state{MA}\country{USA}}
\email{yungew@mit.edu}
\renewcommand{\shortauthors}{Yuxuan Weng and Yunge Wen}

\begin{abstract}
Weather is a complex, continuously changing system in which uncertainty is intrinsic. Physicalizing this uncertainty introduces further variation because computational outputs cannot fully determine material behavior. We distinguish computational uncertainty from material variability and introduce bounded material agency: computation constrains material realization without fixing its exact appearance. We present NephoCodex, a data physicalization system informed by a formative study that constructs five artistic weather states and predicts probability distributions over them. Probability-weighted mappings translate these distributions into material control proposals, while entropy-based regulation, local sensing, and safety constraints bound their execution through mist, airflow, light, and transparent displays. A within-participant study found increased spatial presence and physical demand, while perceived data comprehensibility remained inconclusive after correction. These findings contribute to hybrid data physicalization by showing how variable material expression can be paired with stable digital annotations and how embodied experience can be evaluated separately from data comprehension.
\end{abstract}
\begin{teaserfigure}
\centering
\includegraphics[width=\textwidth,keepaspectratio]{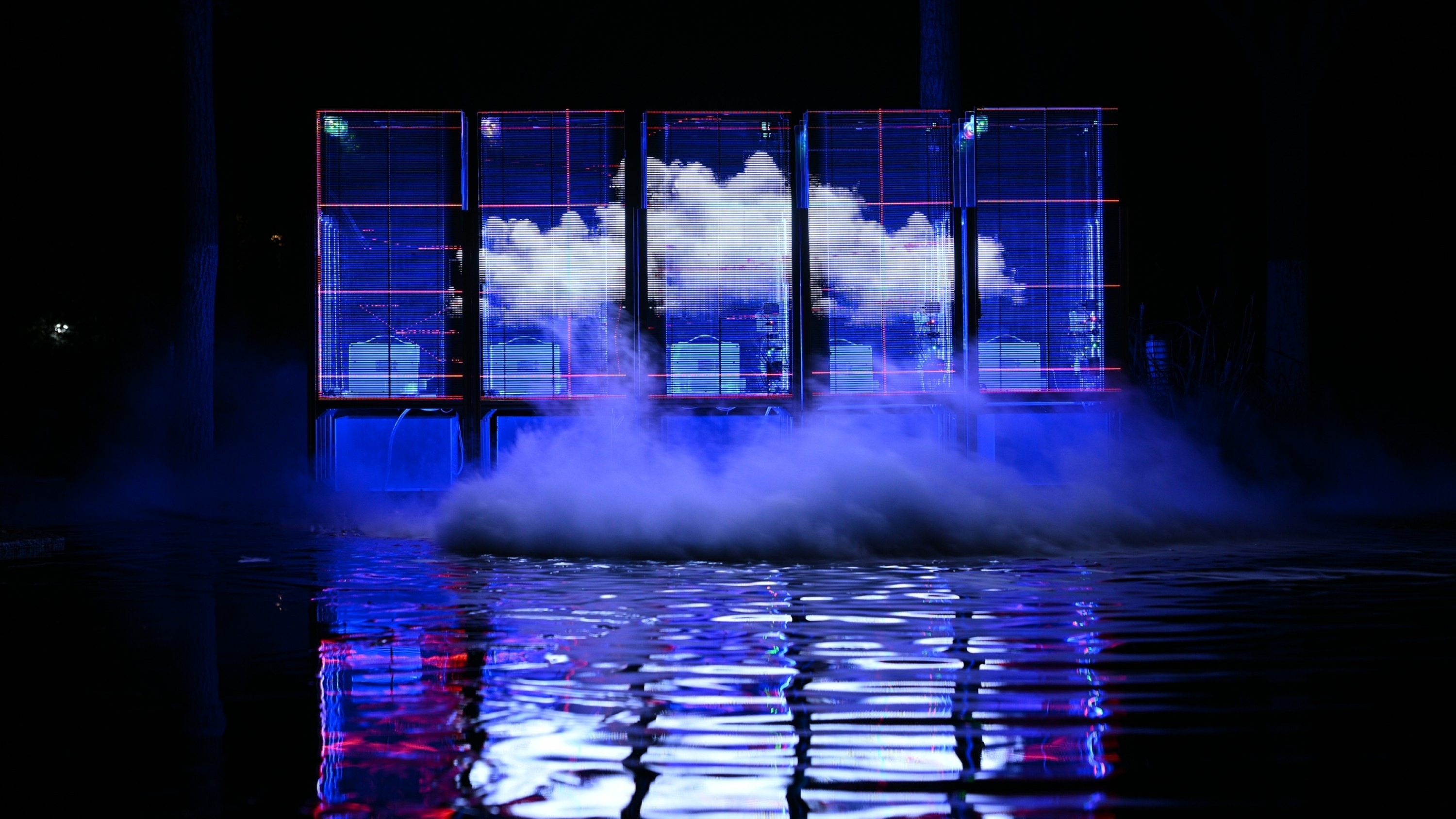}
\caption{NephoCodex maps weather data probabilistically to artistic states, then translates them into proposed controls for mist, airflow, and light. Entropy-based regulation, local sensing, and safety constraints govern execution. Computational uncertainty can thereby influence material variability. We describe this relationship as bounded material agency: computation constrains how a material behaves without fully determining its physical form.}
\label{fig:1}
\Description{Five transparent NephoCodex modules with visible mist, illumination and digital labels.}
\end{teaserfigure}
\maketitle
\section{Introduction}\label{sec:1}

Public-facing weather interfaces often simplify or omit uncertainty in forecasts \cite{hullman2020uncertainty,morss2008uncertainty}. An icon, number, or color gives a compact answer to questions such as whether to carry an umbrella, but may obscure whether the model strongly favors one outcome or assigns nearly equal probabilities to several. Stable symbols make forecasts easy to read and compare. The weather they describe, however, changes continuously: moisture accumulates and dissipates, airflow shifts, and visibility can change unexpectedly. When uncertainty is omitted, a forecast can appear more certain than the model supports.

Tangible interaction and data physicalization encode information through material properties, spatial relationships, and bodily action \cite{hornecker2006tangible,ishii1997tangible,jansen2015physicalization}. More-than-human approaches also question the limits of human-centered design \cite{giaccardi2020morethanhuman}. We ask how probabilistic predictions can guide materials whose behavior varies under explicit control rules, and how this affects viewers' experience and understanding of weather data. To evaluate this relationship, we distinguish presence, cognitive effort, and comprehension. Claims about a display's effectiveness often combine these outcomes, even though immersion can increase without a corresponding improvement in understanding.

We conducted a formative study with six experts in meteorology, data visualization, interaction design, and media art to identify cues that make a physical cloud legible as weather data. Their accounts informed the design, construction, and evaluation of NephoCodex. The system uses near-term weather uncertainty to shape mist, airflow, and light in a transparent enclosure that visitors can enter. A calibrated model reads the preceding day's weather data and predicts probabilities for five artist-defined atmospheric states over the next few hours. The full distribution weights a mixture of five action anchors for atomization, fan, and lighting targets. Concentrated predictions produce clouds with distinct characteristics; dispersed predictions produce clouds that are difficult to assign to a single state. Before these proposals reach the hardware, local humidity, airflow, accumulated water, and safety rules may reduce or override them. Vapor diffusion, ambient air, and visitors' movement further affect the mist, so identical model outputs can produce different clouds. We call this relationship bounded material agency: computation constrains how the cloud forms, while its appearance remains variable. The term does not attribute intention to mist. Calibrated probabilities, safety adjustments, and logged hardware commands let us trace control decisions, although they cannot fully predict the cloud's appearance.

We compared NephoCodex with conventional two-dimensional weather charts in a within-participant study using behavioral observations, paired questionnaires, and open-ended responses. We examined spatial presence, cognitive engagement, and comprehension separately. NephoCodex reliably increased spatial presence and physical demand. Participants approached and circled the installation and waited for state transitions. Self-reported comprehensibility improved descriptively but did not survive correction for multiple comparisons. The results provide stronger evidence for changes in how viewers occupy and move through the space around a weather representation than for changes in their understanding.

Our contributions are:

\begin{itemize}

\item A system that represents weather data through dynamic mist, airflow, and light, and the concept of bounded material agency to describe control over material behavior without requiring exact physical repeatability.

\item A machine learning pipeline that groups continuous weather conditions into five materially distinguishable artistic states and predicts their probabilities, so predictive uncertainty can influence how materials behave within defined controls.

\item Empirical evidence that materializing weather changes spatial and bodily experience: NephoCodex increased spatial presence and physical demand, while perceived comprehensibility remained inconclusive and cognitive engagement did not reliably increase.
\end{itemize}

Two-dimensional weather charts remain better suited to looking up numerical values, checking units, reading trends, and making comparisons. NephoCodex makes weather intensity, formation, movement, and dissipation perceptible through spatial and material change. We propose combining physical and graphical representations to support these complementary tasks.

\begin{table*}[!t]
\caption{Complementary roles of weather charts, mist-based physicalization, and hybrid presentation.}
\label{tab:1}
\centering
\small
\setlength{\tabcolsep}{3pt}
\renewcommand{\arraystretch}{1.22}
\begin{tabular}{@{}>{\raggedright\arraybackslash}p{\dimexpr 0.32000\linewidth-1.333\tabcolsep\relax}>{\raggedright\arraybackslash}p{\dimexpr 0.32000\linewidth-1.333\tabcolsep\relax}>{\raggedright\arraybackslash}p{\dimexpr 0.36000\linewidth-1.333\tabcolsep\relax}@{}}
\toprule
\textbf{2D weather charts} & \textbf{Mist-based physicalization} & \textbf{Hybrid presentation} \\
\midrule
Precise numerical values & Intensity and processes of change & Materials attract attention \\
Units and legends & Space, depth, and duration & Digital information supports interpretation \\
Trends and location comparisons & Bodily movement and waiting & Experiential and analytical roles are combined \\
\bottomrule
\end{tabular}
\end{table*}

\section{Related Work}\label{sec:2}

\subsection{Visualizing Uncertainty in Complex Systems}\label{sec:2.1}

Temperature, humidity, wind, precipitation, and pressure interact nonlinearly across spatial and temporal scales, making weather outcomes uncertain. Weather visualization must make this complexity useful to researchers, professional forecasters, and the public, whose tasks and expertise differ \cite{haase2000meteorology}. Public-facing applications use values, icons, colors, and temporal curves to answer practical questions, retaining or omitting information according to the task. Data processing and interpretation introduce further uncertainty. Making this uncertainty visible requires helping users recognize it and judge how much to trust the representation \cite{sacha2016uncertainty}.

Even familiar displays can be misinterpreted. Morss et al. found that members of the public infer uncertainty from deterministic forecasts and differ in how they interpret expressions such as precipitation probabilities \cite{morss2008uncertainty}. Viewers may also read a hurricane's cone of uncertainty as its impact area or focus excessively on the center line \cite{broad2007cone}. Communicating uncertainty appropriately can improve decisions and sustain trust when forecasts are wrong \cite{joslyn2012uncertainty}. These findings favor clear, distinguishable encodings that retain uncertainty and support action, with visual novelty a secondary concern. Correll and Gleicher compare encodings of means and error \cite{correll2014errorbars}. Hypothetical outcome plots show successive sampled outcomes \cite{hullman2015hop}, while quantile dotplots represent distributions in everyday transit predictions \cite{kay2016transit}. These approaches support retaining uncertainty, but do not establish that physical variability communicates a probability distribution.

Probability calibration concerns how closely confidence corresponds to observed correctness \cite{guo2017calibration}. Using probabilities to guide a representation requires attention to their magnitudes as well as the highest-ranked class. In NephoCodex, calibrated probabilities and predictive entropy guide material control, moving ambiguous predictions toward moderate outputs. Whether viewers interpret those changes as uncertainty requires separate evaluation.

Physikit lets households map sensed environmental data to light, movement, vibration, and airflow \cite{houben2016physikit}. CairnFORM uses expanding, illuminated rings to present forecasts of renewable energy availability in peripheral locations \cite{daniel2019cairnform}. Together, they demonstrate configurable physical outputs for environmental measurements and changing forecasts. NephoCodex examines how probabilities can weight material controls that remain subject to local execution constraints. Presenting forecasts through physical change does not itself establish that viewers can read probabilistic uncertainty from those changes.

Stable two-dimensional symbols support precise lookup and comparison, but have difficulty conveying the uncertainty and continuous formation, movement, and dissipation of weather. NephoCodex examines how control rules translate uncertainty into a changing material representation, and what viewers can interpret from it.

\subsection{Material Agency}\label{sec:2.2}

Tangible interaction connects digital information with physical environments. Ishii and Ullmer's Tangible Bits vision includes graspable objects, interactive surfaces, and ambient media such as light, sound, airflow, and water \cite{ishii1997tangible}. Data physicalization asks how data acquire physical form. Jansen et al. describe opportunities to use three-dimensional space, touch, and bodily movement, alongside challenges in encoding accuracy, updating, interaction, and evaluation \cite{jansen2015physicalization}. Bae et al. organize the design space through context, structure, and interaction, relating understanding to the object, its setting, and how viewers approach it \cite{bae2022tangible}. Viewers must still interpret these spatial and material relationships. Djavaherpour et al. examine the transition from digital design to physical fabrication \cite{djavaherpour2021rendering}. Waldschütz and Hornecker discuss decisions about selecting and staging data for physicalization \cite{waldschutz2020curation}. These accounts inform our explanation of how we construct artistic states and choose actuator anchors.

inFORM uses an actuated pin surface to provide dynamic physical affordances and constraints \cite{follmer2013inform}. HydroMorph controls the shape of a flowing water membrane for display and interaction \cite{nakagaki2016hydromorph}. Both systems demonstrate control over physical form, using different mechanisms for a mechanical surface and a flowing material.

MistForm uses a shape-changing fog screen for interaction with projected content, providing a closely related example of fog as an interactive display medium \cite{tokuda2017mistform}. In NephoCodex, the mist\textquotesingle s density, movement, and dissipation represent weather. Probability-weighted actuator proposals and local execution constraints regulate these properties, while diffusion and surrounding airflow continue to shape the mist.

Ephemeral materials behave differently from stable physicalizations made of wood, plastic, or mechanical components. Döring et al. define ephemeral interfaces as having at least one element deliberately designed to exist for a limited time, often using materials such as water, fire, soap bubbles, or plants \cite{doering2013ephemeral}. Offenhuber's autographic visualizations connect material traces to the conditions that produced them, while retaining the need for scales, labels, context, or comparison \cite{offenhuber2020proxy}. Such materials make time and process visible, but are difficult to reproduce and compare precisely. A stable physicalization can retain its encoded state as viewers inspect it from different angles; an ephemeral material keeps changing. Offenhuber also distinguishes relationships among data, their physical manifestations, and the phenomena they represent \cite{offenhuber2020physicality}. NephoCodex brings these relationships together: a designed mapping makes mist represent remote weather, while the mist also responds physically to local conditions.

Pickering describes scientific practice as a mangle in which human intentions and material resistance shape each other \cite{pickering1995mangle}. More-than-human design questions the limits of centering design on human users \cite{giaccardi2020morethanhuman}. Material agency in HCI need not imply human-like intention. Tholander et al. show how affordances, constraints, and emergent material properties redirect design exploration \cite{tholander2012agency}. Applied to embodied representation, this account draws attention to how material effects change interaction and how designers and systems respond. We use bounded material agency for the relationship between regulated inputs and material effects that remain contingent, without attributing intentions to mist.

Materials also shape the design process. Wiberg proposes a methodology for investigating interaction through materiality \cite{wiberg2014materiality}. Giaccardi and Karana examine how material qualities contribute to experience \cite{giaccardi2015material}, and Material Driven Design connects material exploration with experiential aims \cite{karana2015mdd}. NephoCodex examines how probabilistic model outputs guide actuator proposals, how local rules constrain execution, and how mist varies afterward. Our contribution lies in distinguishing computational uncertainty from material variability within this process. Variable materials and material agency already have precedents in this work.

\subsection{Embodied Presence}\label{sec:2.3}

How viewers encounter a physicalization affects how they understand it. Dourish describes embodied interaction as situated participation: meaning arises through skilled action in an environment as well as abstract decoding \cite{dourish2001action}. Hornecker and Buur connect tangible manipulation, spatial embedding, and social interaction, showing how physical space organizes actions and shared attention \cite{hornecker2006tangible}. Approaching a display, changing viewpoint, or inspecting overlapping layers may help viewers interpret it. Waiting through formation and dissipation or jointly pointing out changes may do so as well. Physecology examines these encounters in terms of a physicalization's setup, interaction mechanisms, and audiences \cite{sauve2022physecology}.

Embedded data representations integrate data with the objects, spaces, or activities they describe, creating trade-offs among visibility, scale, and context \cite{willett2017embedded}. Bodily scale can give abstract quantities a qualitative reference. Data visceralization uses virtual reality to let viewers experience distance, size, and speed at bodily scale, supplementing analytical understanding while risking distortions from scaling \cite{lee2021visceralization}. Physical ambient displays similarly make intensity and duration perceptible through volume, flow, and dissipation. A legible information layer is still needed to relate these sensations to precise values. Chemicals in the Creek connects pollution data with the affected community \cite{perovich2021creek}. It offers an example of situated environmental physicalization, although differences in topic and social setting limit what its outcomes can tell us about environmental learning in NephoCodex.

Sublimate combines actuated physical shapes with virtual graphics and explores transitions between them \cite{leithinger2013sublimate}. This combination of physical and digital information provides a precedent for NephoCodex, which pairs changing mist with a transparent information layer to help viewers relate material behavior to weather values. The usefulness of our pairing still requires evaluation in its own setting; Sublimate\textquotesingle s results cannot establish the effectiveness of NephoCodex's transparent display layer.

We distinguish two forms of embodied effort in this literature. Constitutive actions reveal information: lateral movement exposes layers, waiting reveals transitions, and approaching links material form to annotations. Compensatory actions address obstacles such as occlusion, spatially separated comparison targets, or changes that occur too quickly. Both can increase bodily participation and presence, but only constitutive actions show how embodiment contributes information. Whether greater presence and physical engagement with an uncertain, changing material representation deepen understanding remains underexamined. Controlled studies of physical visualizations assess performance on specified tasks \cite{jansen2013efficiency}, and visualization evaluation frameworks distinguish experience from performance and other goals \cite{lam2012evaluation}. Benefits for uncertainty estimation may also fail to transfer to decision-making \cite{kale2020reasoning}. These studies motivate assessing presence, perceived understanding, and objectively demonstrated interpretation separately.

\section{Formative Study}\label{sec:3}

The formative study asked how viewers without meteorological training could recognize a physical cloud as weather data rather than a stage effect. We examined how explicitly to communicate the mappings from data to materials and how much material variation viewers could accept.

\subsection{Participants}\label{sec:3.1}

We recruited six experts (E1-E6) through interdisciplinary faculty recommendations and online calls. Their backgrounds covered data visualization, meteorology, interaction design, visual communication, media art, and exhibition design (Table~\ref{tab:2}). Participants received the equivalent of USD 10 per hour, prorated to interview duration. The study followed applicable institutional ethics procedures; committee and approval details are withheld for anonymous review.

\subsection{Study Design and Procedure}\label{sec:3.2}

Each individual semi-structured interview lasted approximately 30-40 minutes and was audio-recorded with informed consent. Interviews followed three stages.

Recall. Participants described how they encountered weather: checking applications, reading radar images, viewing animated wind fields, sensing outdoor humidity and airflow, and noticing conflicts between bodily experience and application data. These accounts established how they ordinarily understood weather.

\begin{table*}[!t]
\caption{Expert participants in the formative study.}
\label{tab:2}
\centering
\small
\setlength{\tabcolsep}{3pt}
\renewcommand{\arraystretch}{1.22}
\begin{tabular}{@{}>{\raggedright\arraybackslash}p{\dimexpr 0.08000\linewidth-1.500\tabcolsep\relax}>{\raggedright\arraybackslash}p{\dimexpr 0.17000\linewidth-1.500\tabcolsep\relax}>{\raggedright\arraybackslash}p{\dimexpr 0.12000\linewidth-1.500\tabcolsep\relax}>{\raggedright\arraybackslash}p{\dimexpr 0.63000\linewidth-1.500\tabcolsep\relax}@{}}
\toprule
\textbf{ID} & \textbf{Background} & \textbf{Experience} & \textbf{Professional practice} \\
\midrule
E1 & Data visualization & 10 years & Visual analysis and presentation of complex data, including interactive charts, dashboards, and public-facing data communication. \\
E2 & Media art and exhibition design & 12 years & Digital media art and exhibition spaces, including immersive displays, interactive installations, and multimedia narratives for science exhibitions. \\
E3 & Meteorology & 5 years & Analysis of meteorological observations and interpretation of weather information, including weather-variable displays, forecast communication, and public meteorological education. \\
E4 & Interaction design & 15 years & Digital products and interactive systems, including user research, information architecture, interaction flows, and usability evaluation. \\
E5 & Visual communication & 16 years & Visual information design and communication, including layout, color encoding, graphical symbols, and visual systems across media. \\
E6 & Meteorology & 10 years & Analysis and application of weather and climate information, including meteorological data interpretation, hazard-risk communication, and evaluation of weather services. \\
\bottomrule
\end{tabular}
\end{table*}

\begin{figure*}[!t]
\centering
\resizebox{\linewidth}{!}{\input{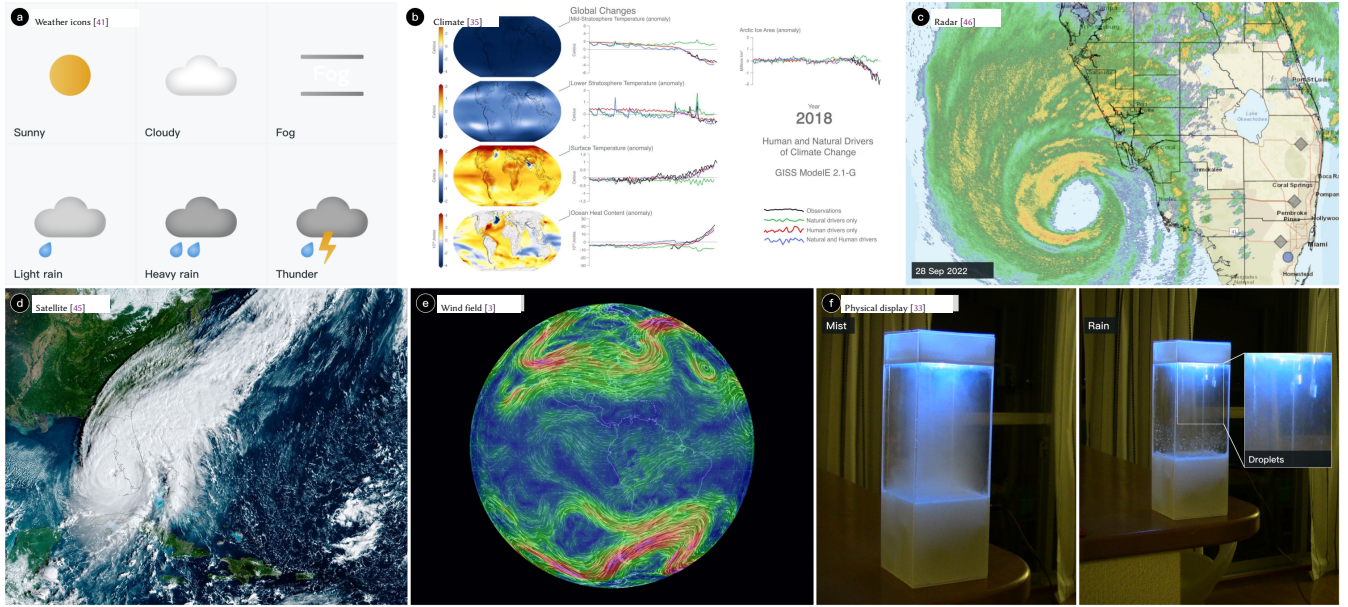}}
\caption{Six forms of weather and climate representation: (a) weather icons \cite{metoffice_weather_symbols}; (b) climate maps and time series \cite{kostis_perkins_climate_drivers_2021}; (c) radar imagery of Hurricane Ian on 28 September 2022 \cite{nhc_ian_radar_2022}; (d) satellite cloud imagery from the same day \cite{nasa_ian_florida_2022}; (e) a static wind-field preview archived in 2013 \cite{beccario_earth_preview_2013}; and (f) mist, rain, and droplet details from OpenTempescope \cite{kawamoto_opentempescope}. These examples support comparison of representational forms. They come from different times and places and do not constitute a matched weather dataset.}
\label{fig:2}
\Description{Six examples of weather representation: icons, climate trends, radar, satellite cloud imagery, wind flow and a physical weather display.}
\end{figure*}

Compare. Participants compared icons, static charts, animations, satellite imagery, dynamic wind maps, and physical representations in terms of category recognition, trend interpretation, numerical lookup, and spatial experience. Figure~\ref{fig:2} shows the representational forms used in this activity.

Sources for Figure~\ref{fig:2}: (b) NASA's Scientific Visualization Studio; (c) NWS/NHC, archived on Wikimedia Commons; (d) Joshua Stevens, using GOES-16 imagery courtesy of NOAA/NESDIS; (e) Cameron Beccario; and (f) Ken Kawamoto.

Construct. Participants mapped temperature, humidity, wind, cloud cover, and precipitation to possible mist amounts, densities, directions, speeds, lighting effects, and textual annotations. They proposed physical effects and identified mappings that might invite conflicting interpretations or confuse viewers.

We analyzed recordings, field notes, sketches, sticky notes, and mapping selections in two rounds. First, we retained descriptions of familiar weather practices, confusing visual encodings, proposed mappings, and tolerance for variation. We then compared descriptions across participants and grouped them around three recurring concerns: existing weather experiences, the intelligibility of data-to-material mappings, and acceptable material variation. These concerns informed the design implications below.

\subsection{Findings}\label{sec:3.3}

Participants distinguished recognizing a weather category quickly from perceiving weather as a process. E1 noted that conventional icons quickly convey sunny, cloudy, or rainy conditions, while dynamic wind maps show continuity between states. This suggests that familiar categories gain efficiency by omitting some of the continuity needed to understand weather as it unfolds.

E2 found mist appealing because fans and space jointly produce it as an event, with variation contributing to its artistic quality. To use that variation as information, viewers needed to know its source. E2 emphasized distinguishing weather-driven changes from those caused by “the air conditioning in the room.” A data mapping therefore needs to separate intended material variation from environmental noise.

E4 valued ambient media that viewers can perceive without focused attention. Having to infer six variables from each cloud, however, would make this a demanding decoding task. The benefit of peripheral perception would be lost if interpreting an unexplained environment simply replaced checking an application.

\subsection{Design Implications}\label{sec:3.4}

The analysis yielded three directions for making physical clouds recognizable and interpretable as weather data.

D1. Begin with familiar weather experiences and show continuous formation, movement, and dissipation. Material changes should evoke recognizable processes such as clearing, thickening, accumulating, or becoming unstable. NephoCodex uses five artistic states whose probability-weighted mappings allow gradual transitions while retaining familiar weather cues.

D2. Keep directional mappings consistent and show data sources and timing. Viewers should be able to relate material changes to the data and check where and when those data originated. NephoCodex derives atomization, fan, and lighting parameters from predicted state probabilities and displays numerical values, timestamps, and sources.

D3. Name states and display key values and mapping cues. The accompanying information should explain what each state represents and what viewers can infer from it. NephoCodex's transparent displays identify the states as artistic interpretations, show data and legends, and make clear that the installation is not a warning system.

\section{NephoCodex}\label{sec:4}

\begin{figure*}[!t]
\centering
\includegraphics[width=\linewidth,keepaspectratio]{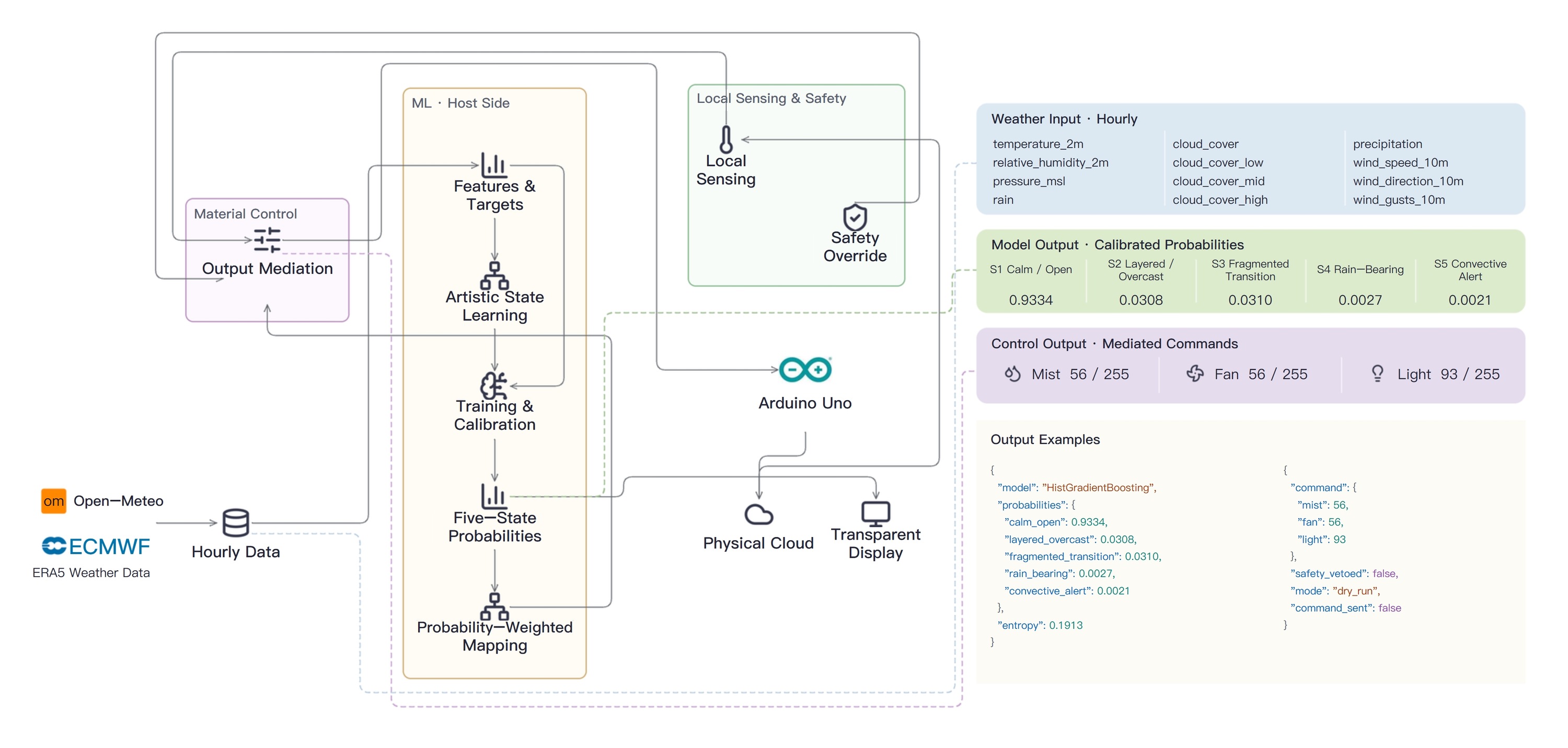}
\caption{NephoCodex system overview. Weather data pass through state prediction, probability mapping, and local adjustment to generate atomization, fan, and lighting commands. The digital display layer provides states, values, timestamps, and sources.}
\label{fig:3}
\Description{System pipeline linking weather inputs, state probabilities, local controls, mist, airflow, lighting and digital labels.}
\end{figure*}

NephoCodex represents weather through mist, airflow, and light. It constructs five artistic states from historical weather data, then uses current and past records to estimate their probabilities for the next three hours. A controller converts the distribution into proposed material parameters and issues actuator commands subject to local conditions and constraints. Transparent displays show state names, key values, timestamps, and sources, allowing viewers to consult the data behind the mist. Figure~\ref{fig:3} shows the system.

The design uses familiar weather cues (D1), makes the conversion from data to control parameters explicit (D2), and explains the states and their artistic status through digital annotations (D3). It controls atomization, fans, and lighting; the cloud's morphology emerges as these inputs interact with the materials and environment.

\subsection{Weather Prediction Model}\label{sec:4.1}

The prediction pipeline links continuous weather records to NephoCodex's material controls. It first groups future weather conditions into five states. Designers interpret these groups as artistic states and assign morphological references and material control anchors. A supervised model then predicts a distribution over the states. This lets computational uncertainty influence the proposed material expression while leaving its exact appearance unspecified.

\subsubsection{Dataset Curation}\label{sec:4.1.1}

We developed the model using hourly ERA5 reanalysis data \cite{hersbach2020era5} obtained through the Open-Meteo Historical Weather API \cite{openmeteo2026historical}. The records cover 1 January 2020 to 31 December 2025 and include temperature, humidity, cloud cover, precipitation, wind speed, gusts, and pressure. We describe the location as a tropical city and withhold its precise coordinates for anonymous review.

The original archive contained 52,608 hourly records. Removing duplicate or incomplete records and requiring 24 hours of historical features plus a three-hour target window left 52,581 usable samples. We retained the data request information as provenance metadata.

At each prediction time t, the model uses records timestamped no later than t, including current values and up to 24 hours of historical features, to predict the state for the t + 1 to t + 3 hour window. Each target describes the whole window, rather than three independent hourly labels. ERA5 supplies gridded reanalysis estimates, not measurements from the installation's sensors. Because historical reanalysis differs from information available in real time, this model supports the representational pipeline rather than operational forecasting. Remote records supply weather content; local sensors constrain how the installation expresses it.

\subsubsection{Clustering}\label{sec:4.1.2}

We fitted k-means only on the training period, grouping future one-to-three-hour weather conditions by mean and maximum cloud cover, total and maximum precipitation, mean humidity, maximum wind speed and gusts, and pressure change \cite{macqueen1967classification}. These eight descriptors construct discrete prediction targets from continuous conditions. They describe the future window and are unavailable as inputs at prediction time. Table~\ref{tab:A1} and Table~\ref{tab:A2} in the appendix list the descriptors and cluster centers.

We compared K = 4 through K = 7 on 10,000 training samples. Their silhouette scores \cite{rousseeuw1987silhouettes} were .304, .307, .232, and .231, respectively. K = 5 scored highest, but its .003 advantage over K = 4 does not establish a uniquely optimal taxonomy. We also considered what the installation could express: thinness and persistent coverage, fragmentation, heavier precipitation, and disturbance. Further distinctions would be useful only if atomization, airflow, and lighting could convey them. The five classes reflect these material design choices as well as the data groups; they do not specify the number of categories in natural weather.

\begin{figure*}[!t]
\centering
\includegraphics[width=\linewidth,keepaspectratio]{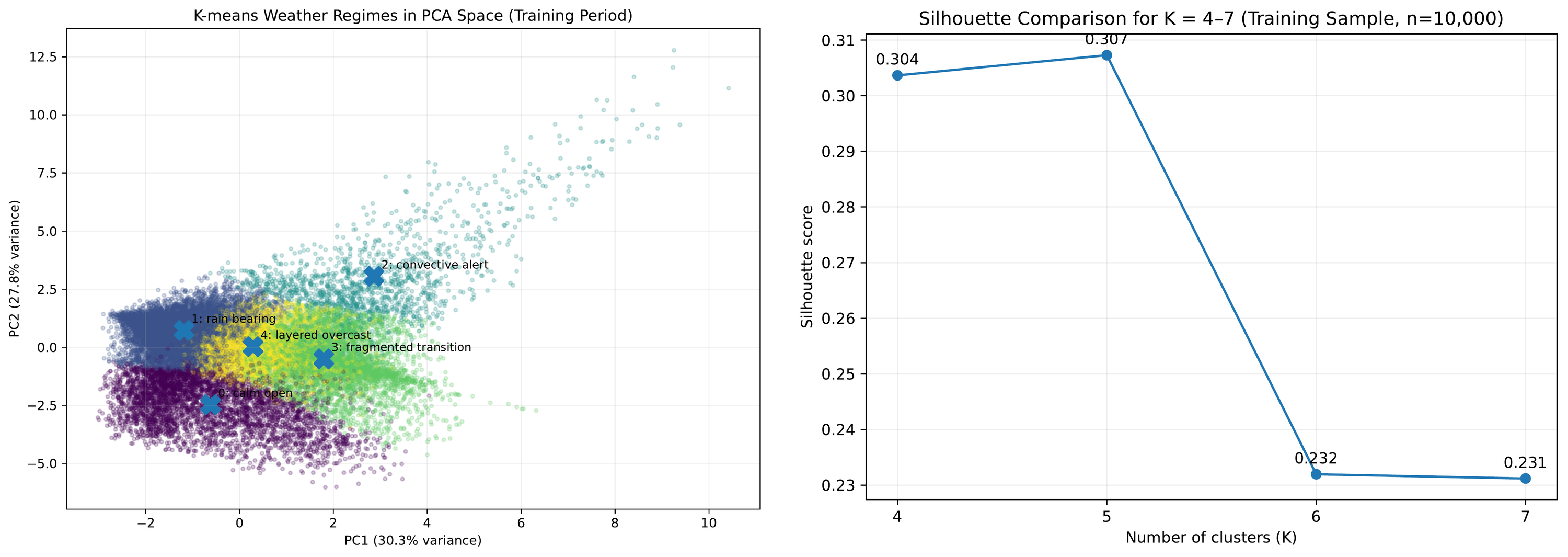}
\caption{Clustering analysis for weather state construction. Left: training-period data projected onto the first two principal components, jointly explaining 58.1\% of the variance. Right: silhouette scores for four through seven classes, compared using 10,000 training samples.}
\label{fig:4}
\Description{PCA projection of weather samples and silhouette scores for four to seven clusters.}
\end{figure*}

Figure~\ref{fig:4} projects the training data onto the first two principal components, which jointly explain 58.1\% of the variance, and shows silhouette scores for the candidate values of K. These analyses describe the state space we constructed. They do not independently validate the classes or establish their stability.

\subsubsection{Five Artistic States}\label{sec:4.1.3}

We interpreted the clusters' meteorological characteristics as five artistic states: calm/open, layered/overcast, fragmented transition, rain-bearing, and convective alert. Designers assigned these names; they are not independently annotated meteorological ground truth. We then used the morphology of cirrostratus, stratocumulus, altocumulus, nimbostratus, and cumulonimbus to guide coverage, density, separation, and disturbance.

\begin{figure}[!htbp]
\centering
\includegraphics[width=\linewidth,keepaspectratio]{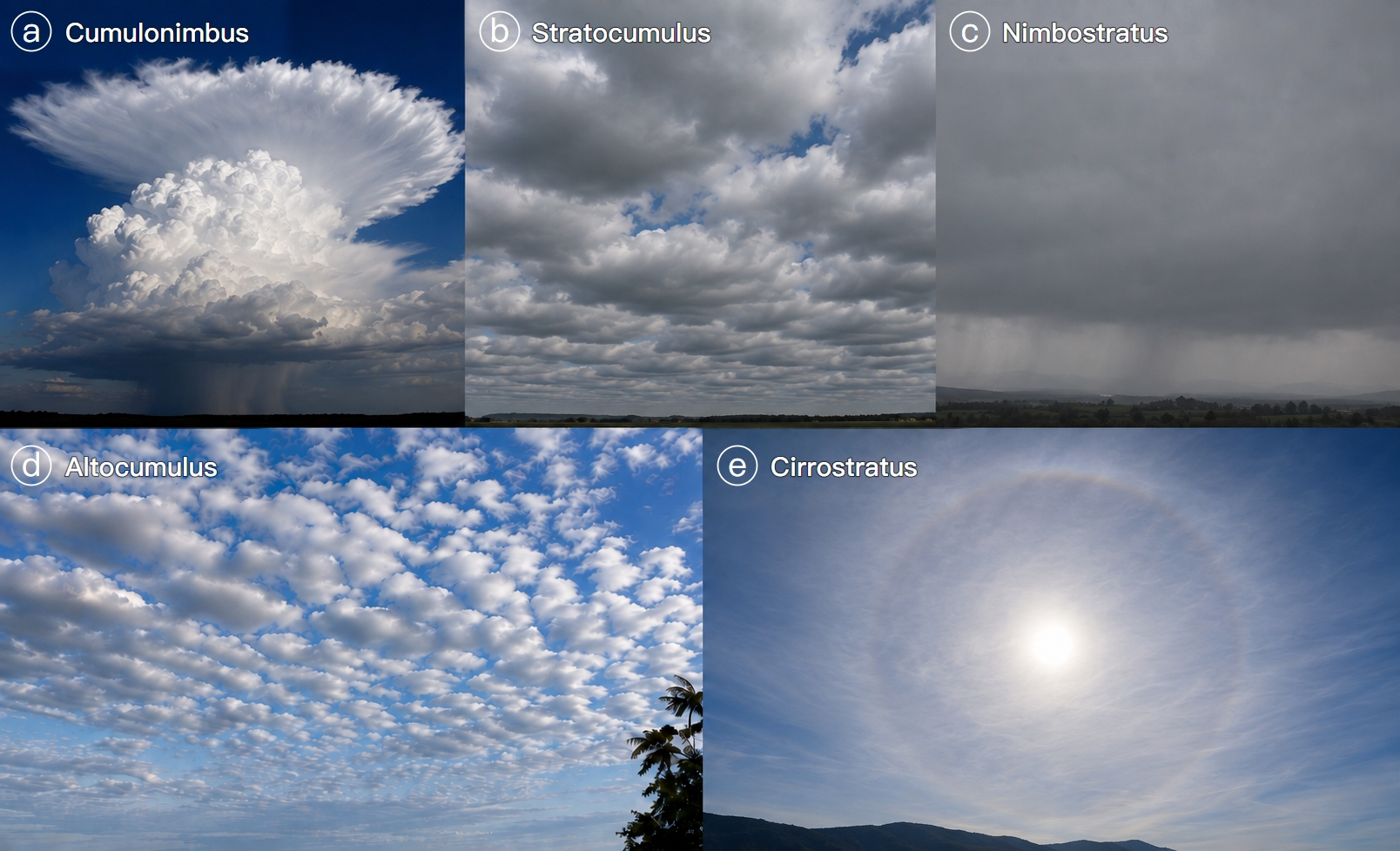}
\caption{Five natural cloud forms used as material-design references: (a) cumulonimbus, (b) stratocumulus, (c) nimbostratus, (d) altocumulus, and (e) cirrostratus. Cloud genera follow the WMO International Cloud Atlas \cite{wmo2017atlas}. These photographs illustrate morphology rather than the system's classification outputs.}
\label{fig:5}
\Description{Five reference cloud photographs: cumulonimbus, stratocumulus, nimbostratus, altocumulus and cirrostratus.}
\end{figure}

Cirrostratus's thin, veil-like form inspired light, open layers. Stratocumulus suggested persistent horizontal coverage through its cellular and layered structure, while altocumulus suggested fragmentation and visible separation. The heavy appearance of nimbostratus informed dense accumulation with subdued lighting; the vertical development of cumulonimbus informed stronger disturbance. We used these associations to guide material design, without assuming one-to-one correspondences between weather variables and natural cloud genera. Figure~\ref{fig:5} shows the five morphological references.

\begin{figure*}[!t]
\centering
\includegraphics[width=\linewidth,keepaspectratio]{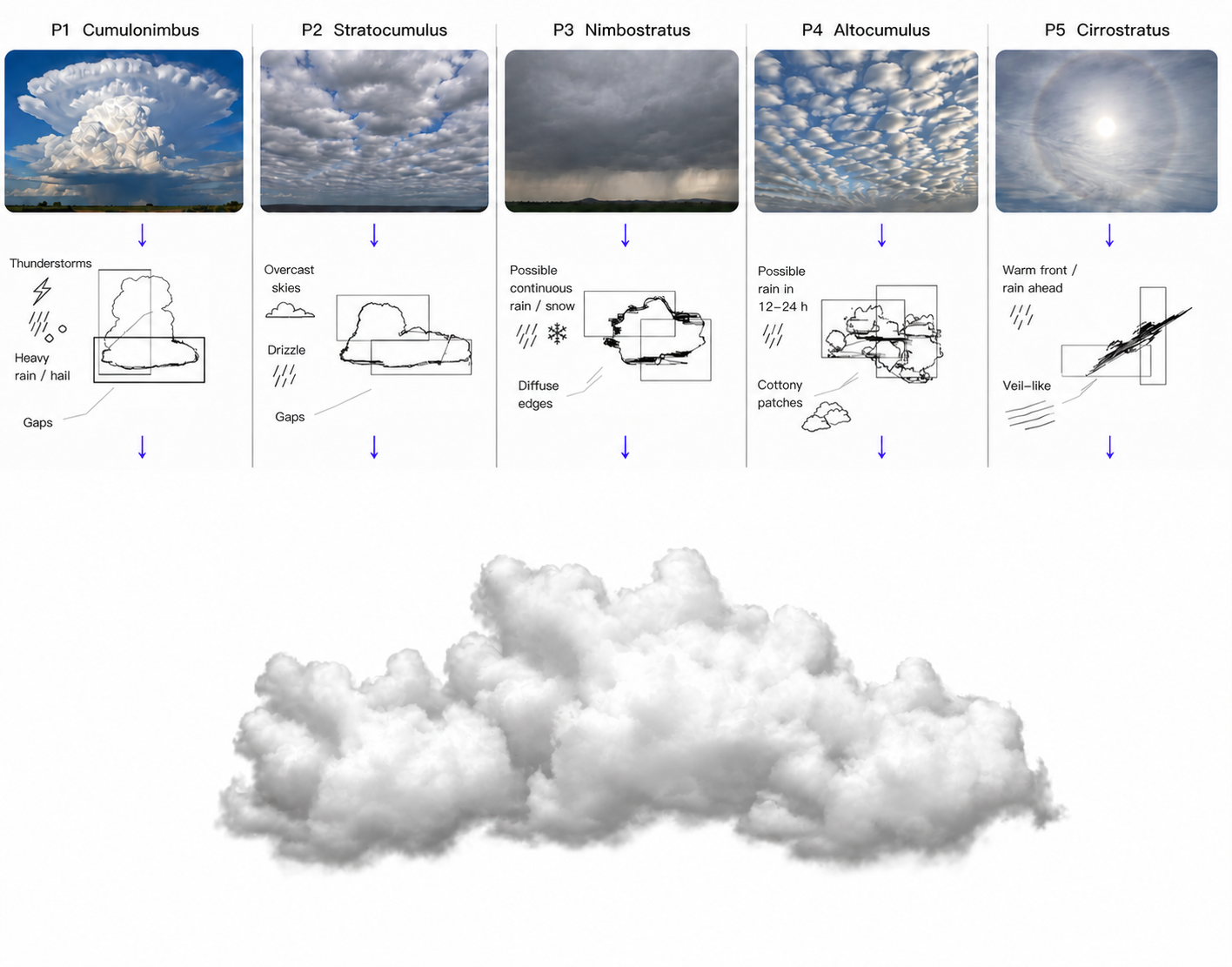}
\caption{Morphological references and material expressions for the five artistic states. Natural cloud forms provide a vocabulary of coverage, density, and disturbance. The correspondences represent design mappings, not recognition of natural cloud genera.}
\label{fig:6}
\Description{Design correspondence between five artistic weather states, cloud morphology and material expression.}
\end{figure*}

Clustering groups weather conditions. Designers then assign names, morphological references, and actuator anchors in a separate mapping step. Cluster labels contain no cloud shapes, so they cannot by themselves determine atomization, fan, or lighting parameters. Figure~\ref{fig:6} shows how we mapped natural cloud morphology to material expression.

The WMO International Cloud Atlas describes cloud genera by observable characteristics \cite{wmo2017atlas}. We use these forms as references for material expression.

NephoCodex uses cloud morphology as a design reference. It neither identifies natural cloud genera nor reproduces their altitude, microphysics, or atmospheric scale. Convective alert denotes heightened artistic intensity, not thunderstorm detection or a public warning. The digital layer labels all five states as artistic interpretations. Their control anchors (Table~\ref{tab:3}) specify positions in a software control space, not measured physical quantities. We have not independently validated whether viewers can distinguish the intended expressions.

\begin{table*}[!t]
\caption{Control anchors and intended expressions for the five artistic states. Values are percentages of the software control range, ordered as atomization, fan, and lighting.}
\label{tab:3}
\centering
\small
\setlength{\tabcolsep}{3pt}
\renewcommand{\arraystretch}{1.22}
\begin{tabular}{@{}>{\raggedright\arraybackslash}p{\dimexpr 0.18000\linewidth-1.500\tabcolsep\relax}>{\raggedright\arraybackslash}p{\dimexpr 0.13000\linewidth-1.500\tabcolsep\relax}>{\raggedright\arraybackslash}p{\dimexpr 0.24000\linewidth-1.500\tabcolsep\relax}>{\raggedright\arraybackslash}p{\dimexpr 0.45000\linewidth-1.500\tabcolsep\relax}@{}}
\toprule
\textbf{State} & \textbf{Morphology} & \textbf{Atomization / fan / light (\%)} & \textbf{Intended material expression} \\
\midrule
Calm/open & Cirrostratus & 20 / 20 / 35 & Thin layers with slow movement \\
Layered/overcast & Stratocumulus & 42 / 30 / 50 & Horizontal spread and persistent coverage \\
Fragmented transition & Altocumulus & 55 / 62 / 65 & Separated cloud groups appearing and moving intermittently \\
Rain-bearing & Nimbostratus & 80 / 58 / 42 & Dense accumulation with subdued lighting \\
Convective alert & Cumulonimbus & 96 / 95 / 82 & Strong atomization and airflow disturbance; artistic intensity only \\
\bottomrule
\end{tabular}
\end{table*}

\subsubsection{Prediction Model}\label{sec:4.1.4}

We trained a supervised predictor on the constructed labels using 105 features from current and historical weather records. These features include current conditions, lagged values, rolling statistics, recent changes, precipitation history, and temporal encodings. They summarize the weather's current state, recent change, temporal variability, and periodicity.

We compared random forests with histogram-based gradient boosting \cite{friedman2001gradient,pedregosa2011scikit} by validation log loss \cite{gneiting2007scoring} and selected the latter. We then applied temperature scaling \cite{guo2017calibration} on a separate calibration period (T = 1.40). For each future window, the model predicts a probability distribution over the five states.

The probabilities weight the five material control anchors directly. A concentrated distribution moves the proposal closer to one anchor; a dispersed distribution combines contributions from several states. Temperature scaling adjusts concentration while preserving class ranking, so it cannot correct an incorrect highest-probability class. Section~\ref{sec:5.2} evaluates test performance and probability calibration.

The distribution specifies proposed material controls, leaving the final physical state undetermined. Environmental sensing and safety regulation can modify actuator values before execution, and airflow, droplets, and surrounding conditions continue to affect the mist afterward. Prediction, regulated execution, and the resulting physical form are therefore separate stages.

\subsection{Hardware and Physical Constraints}\label{sec:4.2}

The controller converts predicted states into actuator commands through probability weighting, uncertainty regulation, and local constraints (Figure~\ref{fig:3}). Computational uncertainty influences the proposed material parameters. Environmental conditions and safety rules can then modify or override them before execution.

\subsubsection{Probability Weighting and Entropy-Based Regulation}\label{sec:4.2.1}

Let \(p_{t,k}\) denote the probability of state k at time t, and let \(v_k\) be its material anchor, whose three components specify atomization, fan, and lighting levels. The initial control proposal is

\begin{equation}
a_t=\sum_{k=1}^{5}p_{t,k}v_k
\label{eq:1}
\end{equation}

As probability concentrates on one state, the proposal approaches that state's anchor. When probability is distributed across states, it lies between anchors, with every state of nonzero probability contributing. If the two most probable classes exchange rank, the proposal need not switch an entire anchor vector.

Probability weighting makes the control proposal change continuously as probabilities change. Constraint rules and physical responses may still introduce discontinuities, so the visible mist need not change smoothly. Different distributions can also yield the same weighted actuator vector. Some distributional information is therefore lost in the physical output, even though every class probability contributes to control; viewers cannot recover the full distribution from the mist.

The system uses normalized predictive entropy to describe dispersion in the probability distribution:

\begin{equation}
U_t=-\frac{\sum_{k=1}^{5}p_{t,k}\log p_{t,k}}{\log 5}
\label{eq:2}
\end{equation}

When \(U_t\) \(\geq\) .75, the controller moves the proposal toward a preset moderate-output center (35, 30, 45), with a maximum mixing proportion of .65. We set this threshold and mixing limit empirically for the prototype to restrict extreme outputs when predictions are dispersed. Their suitability as general-purpose reliability thresholds has not been validated.

Entropy measures how probabilities are distributed across the five states. It neither distinguishes sources of uncertainty nor measures physical variation in mist. Probability weighting and entropy regulation allow computational uncertainty to affect material parameters within designed bounds. Separate evaluation is needed to determine whether viewers perceive or understand that uncertainty.

\subsubsection{Local Adjustment and Constraint Rules}\label{sec:4.2.2}

Local humidity, airflow, and fault states give the controller information about operating conditions absent from the model inputs. These conditions can modify or override the proposed material parameters before execution.

When local relative humidity is at least 88\% but below 95\%, the atomization proposal is halved; at or above 95\%, atomization output is set to zero. When local airflow reaches the configured upper limit, the fan proposal is reduced by 30\%. Accumulated water, serial communication failures, and emergency stops have higher priority and set all outputs to zero.

After these checks, the controller converts percentage targets into commands in its 0-255 range. The commands specify actuator inputs, not measured mist amounts, wind speeds, or light intensities. They constrain how the mist is produced while leaving its exact appearance variable.

The system logs state probabilities, predictive entropy, initial proposals, adjusted targets, and final commands. The logs show how each command was produced and whether probability weighting, local adjustment, or a constraint override caused a change. They cannot fully explain the cloud's appearance: vapor diffusion, enclosure geometry, ambient airflow, and visitor movement also affect the mist after execution.

\subsubsection{Physical Implementation and Information Presentation}\label{sec:4.2.3}

\begin{figure*}[!t]
\centering
\includegraphics[width=\linewidth,keepaspectratio]{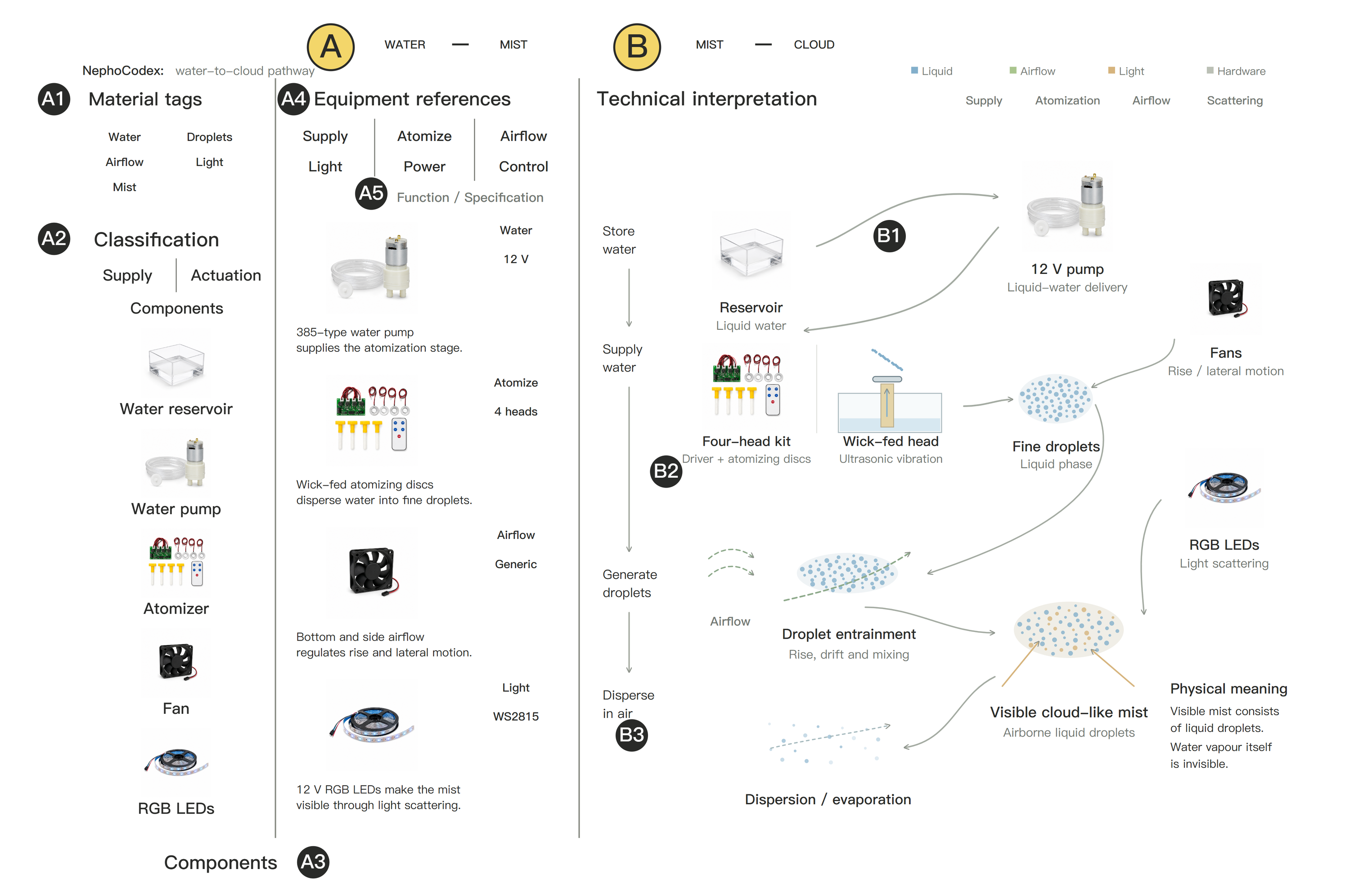}
\caption{Material-generation pathway from water to mist. Water supply, ultrasonic atomization, airflow transport, and illumination jointly produce visible cloud-like mist. The diagram distinguishes materials, equipment, and their physical roles.}
\label{fig:7}
\Description{Water-to-mist pathway showing water supply, ultrasonic atomization, airflow transport and illumination.}
\end{figure*}

\begin{figure*}[!t]
\centering
\includegraphics[width=\linewidth,keepaspectratio]{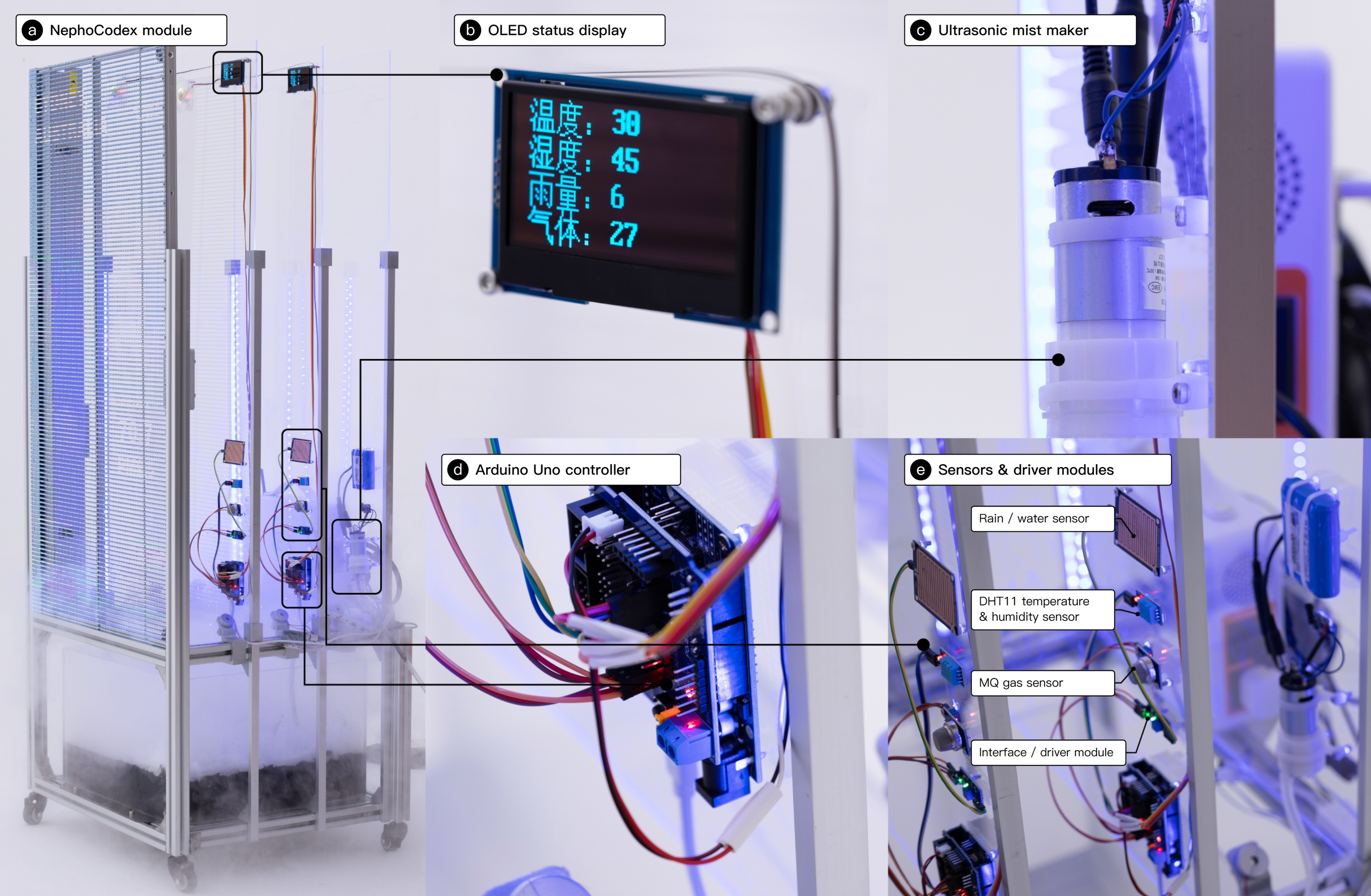}
\caption{NephoCodex modules and hardware, including displays, atomization and water circulation, fan actuators, and control electronics separated from the wet zone.}
\label{fig:8}
\Description{Annotated photographs of modules, display, water circulation, fans and control electronics.}
\end{figure*}

Each module uses ultrasonic atomizers to generate cloud-like mist from a recirculating water supply. Bottom and side fans regulate its movement, and lighting shapes its visibility. Sensors for local temperature and humidity, airflow, and accumulated water supply the inputs used by the constraint rules. Figure~\ref{fig:7} shows the material-generation pathway, and Figure~\ref{fig:8} shows the modules and hardware.

The host sends atomization, fan, and lighting targets to the controller and receives acknowledgments and local sensor states. Aluminum profiles and transparent acrylic panels enclose the observable flow space. The control electronics sit apart from the reservoir and wet zones.

Transparent displays behind the mist place weather values, state names, and material expression in the same viewing direction. They show timestamps, sources, and mapping explanations and identify the states as artistic interpretations. We limit information density and use high contrast for key labels to reduce reading difficulties when dense mist obscures text. The mist conveys process and change, while the digital layer supplies explicit values and helps viewers interpret them. The user study examines how viewers read and interpret this combination.

\section{Technical Evaluation}\label{sec:5}

We evaluated whether the model's probabilistic outputs support the data-to-command pipeline: how well it anticipates the constructed states, how closely its probabilities correspond to observations, and what its software records establish about material control.

We analyzed historical state predictions and compared reliability curves before and after temperature scaling. These analyses assess prediction, the interpretation of probabilities, and software output. They do not directly test the accuracy of the physical cloud, viewers' understanding of uncertainty, or the operational reliability of the complete installation.

\subsection{Evaluation Setup}\label{sec:5.1}

The model uses features available up to time t to estimate the cluster state for the future t +1 to t +3 hour window. Target labels follow Section~\ref{sec:4.1.2}. Agreement with these labels measures how well the model anticipates the constructed data groups. It does not validate the model against independent meteorological ground truth.

We divided the 52,581 usable samples chronologically: the first 60\% for training, the next 15\% for probability calibration, the next 10\% for validation and model selection, and the final 15\% for testing. We fitted k-means on the training period, selected the classifier by validation log loss, and estimated temperature on the calibration period. The test set contains 7,888 hourly predictions from the final historical period at the same location.

We calculated accuracy, balanced accuracy, macro-averaged F1, and class-specific recall from the test confusion matrix. Reliability diagrams compare predicted confidence with observed accuracy before and after temperature scaling. Probability quality matters for control because NephoCodex weights its outputs using the full distribution. The 7,888 samples share historical information and overlapping future windows. They describe performance during this period, not an equivalent number of independent weather events or evidence of generalization across locations.

\subsection{State Prediction and Probabilistic Material Control}\label{sec:5.2}

\begin{figure}[!htbp]
\centering
\includegraphics[width=\linewidth,keepaspectratio]{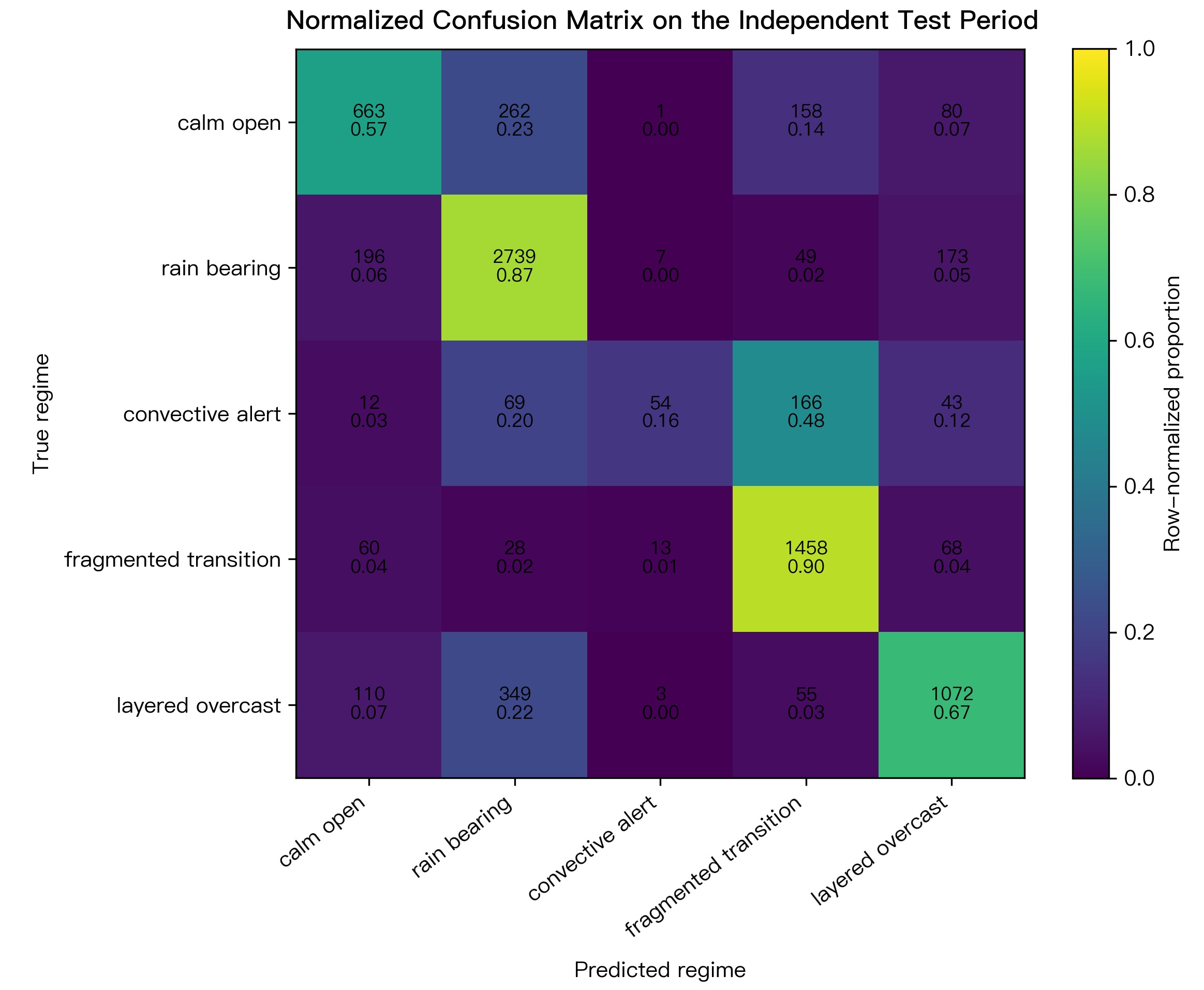}
\caption{State-prediction confusion matrix for the temporal holdout test set. Rows indicate target cluster states and columns indicate predictions. Each cell shows its count above the row-normalized proportion. The matrix contains 7,888 test samples. Target classes are the states constructed in this study, not independent meteorological ground truth.}
\label{fig:9}
\Description{Five-class confusion matrix for 7,888 held-out samples with counts and row-normalized proportions.}
\end{figure}

The model correctly classified 5,986 of the 7,888 test samples (Figure~\ref{fig:9}), yielding an accuracy of .759. Balanced accuracy was .633 and macro-averaged F1 was .645. Performance varied across the five states. Convective alert accounted for 4.4\% of test samples and had a recall of .157 (54/344). The model's usefulness for material control therefore depends on the class. The confusion matrix alone does not explain these differences.

The controller mixes all five action anchors, so probabilities assigned to winning and nonwinning states both affect its actuator proposals. Interpreting these probabilities matters for the physical representation as well as for evaluating the classifier.

\begin{figure}[!htbp]
\centering
\includegraphics[width=\linewidth,keepaspectratio]{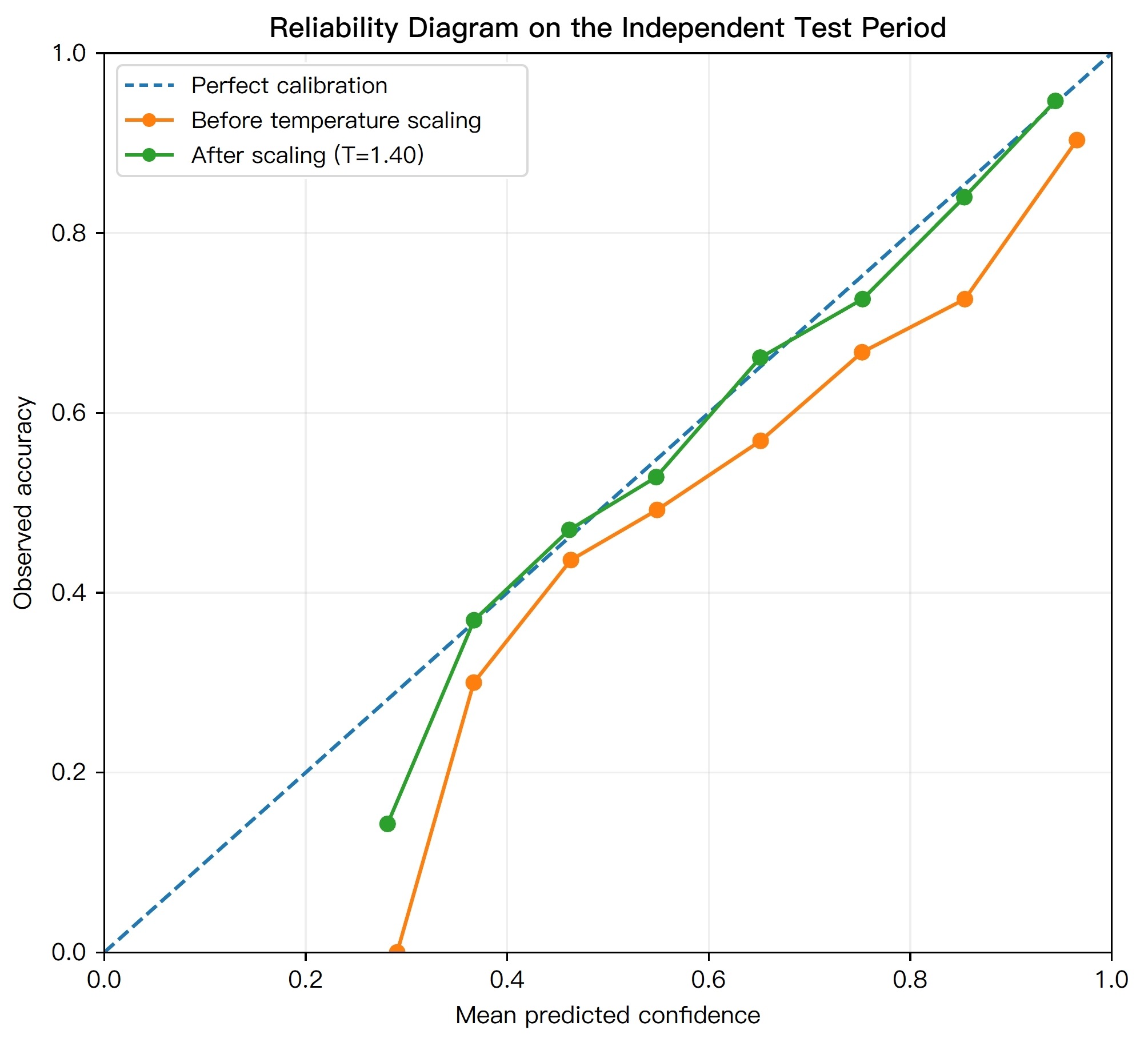}
\caption{Reliability diagrams before and after temperature scaling. The horizontal axis shows mean predicted confidence within each bin; the vertical axis shows observed accuracy. The dashed line indicates ideal calibration. Orange represents the unscaled model, and green represents scaling with \(T=1.40\).}
\label{fig:10}
\Description{Reliability curves before and after temperature scaling with T equal to 1.40.}
\end{figure}

Figure~\ref{fig:10} compares binned mean confidence with observed accuracy before and after temperature scaling. Before scaling, mean confidence exceeds accuracy in most displayed intervals. After scaling, most intervals lie closer to the ideal diagonal, indicating closer correspondence between confidence and accuracy, although deviations remain at low confidence. The diagrams assess overall confidence rather than each class probability, so they provide only partial evidence about the full distributions used in control.

Because nonwinning states contribute to the weighted mixture, a classification error does not translate directly into a discrete material-state error. This mixing does not repair inaccurate probabilities. Entropy regulation, environmental sensing, and safety rules likewise constrain execution without establishing whether the prediction was correct.

The model anticipates the constructed states unevenly across classes, and temperature scaling improves the displayed correspondence between confidence and accuracy. These results support using the distribution to propose material controls. They leave open whether every class probability is well calibrated, whether each distribution produces a unique cloud appearance, and whether the mixtures communicate states or uncertainty accurately.

\section{User Study}\label{sec:6}

\subsection{Participants}\label{sec:6.1}

Thirteen participants (P1-P13; nine women and four men) completed the study. They were 21 to 46 years old. Nine studied interaction design or art-related subjects; the others included university administrative staff, weather enthusiasts, and participants from other backgrounds. We recruited participants online and paid USD 10 per hour.

\subsection{Conditions}\label{sec:6.2}

We conducted a within-participant comparison of two conditions:

\begin{itemize}

\item Weather Charts: two-dimensional weather visualizations displayed on a screen.

\item NephoCodex: the physical cloud installation.
\end{itemize}

\begin{figure*}[!t]
\centering
\includegraphics[width=\linewidth,keepaspectratio]{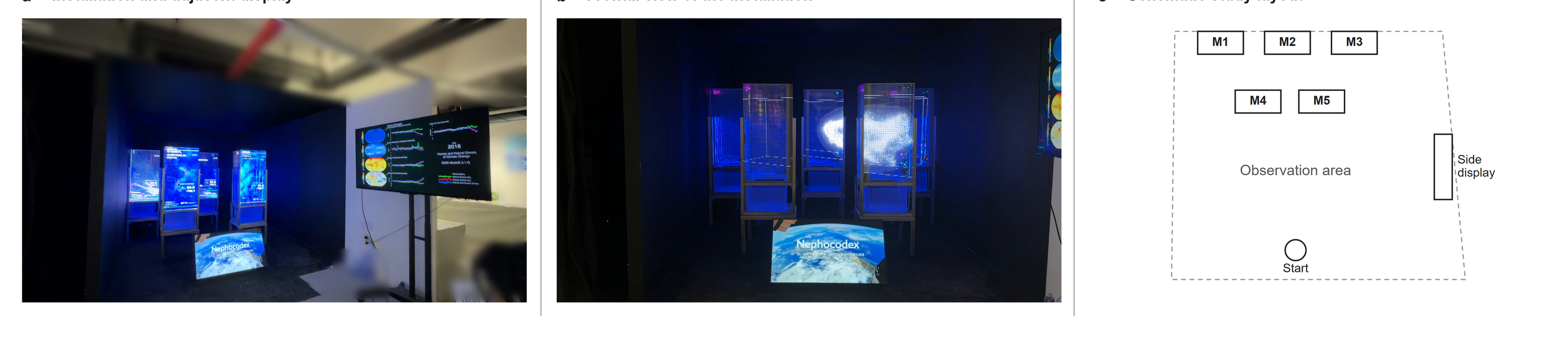}
\caption{Study installation and observation space: (a) the installation and adjacent screen, (b) the front of the installation, and (c) a schematic layout of the modules, observation area, and starting point.}
\label{fig:11}
\Description{Installation photographs and schematic observation layout.}
\end{figure*}

Both conditions showed the same location, period, and weather data to control for differences in the underlying forecast. The study took place on the same day at an outdoor waterfront site; its exact date and location are withheld for anonymous review. Figure~\ref{fig:11} shows the installation and observation space.

\subsection{Procedure}\label{sec:6.3}

The study received institutional ethics approval. We briefed all participants on the procedure and both visualization conditions, and obtained their written informed consent.

We used two presentation orders to control for sequence effects: ABBA (5 participants) and BAAB (8 participants), with A representing Weather Charts and B representing NephoCodex. Participants first completed a 90-second eyes-closed resting baseline. They then experienced each condition for 125 seconds, rested with eyes closed for 30 seconds, and experienced both conditions again in reverse order for 120 seconds each. Each participant therefore encountered each condition twice, totaling 245 seconds per condition. The full experimental sequence lasted 610 seconds.

After the tasks, participants gave one overall rating for each condition and took part in a 5-10-minute open-ended interview. Each session lasted approximately 30 minutes, including instructions, questionnaires, and the interview.

\subsection{Measures}\label{sec:6.4}

We used behavioral observations, self-report questionnaires, and open-ended responses to examine participants' experiences in both conditions.

\subsubsection{Behavioral Observation}\label{sec:6.4.1}

Two cameras, behind and to the rear right of each participant, recorded movement, viewing behavior, and interaction during the tasks. We coded all task-phase recordings from both conditions, excluding baseline and rest periods. Before coding independently, we divided the recordings into shared segments with predefined temporal boundaries. Both researchers completed joint training and pilot coding, then independently labeled each segment's interaction role from its observable action, visible target, and surrounding sequence. We paired their labels by segment to assess agreement; the shared boundaries avoided having to match independently detected events in time. Cohen's \(\kappa\) was .83 for interaction-role labels \cite{cohen1960agreement}. The researchers then discussed disagreements, referring back to the recordings and field notes. This agreement measures how consistently they interpreted predefined segments. It does not assess event segmentation or independently verify participants' intentions.

\begin{table*}[!t]
\caption{Event-based behavioral codebook and potential interaction roles.}
\label{tab:4}
\centering
\small
\setlength{\tabcolsep}{3pt}
\renewcommand{\arraystretch}{1.22}
\begin{tabular}{@{}>{\raggedright\arraybackslash}p{\dimexpr 0.18000\linewidth-1.333\tabcolsep\relax}>{\raggedright\arraybackslash}p{\dimexpr 0.42000\linewidth-1.333\tabcolsep\relax}>{\raggedright\arraybackslash}p{\dimexpr 0.40000\linewidth-1.333\tabcolsep\relax}@{}}
\toprule
\textbf{Code} & \textbf{Operational definition} & \textbf{Potential interaction role} \\
\midrule
B1 Approach & Deliberately reduce distance to a module or display & Inspect values, annotations, or fine mist structure \\
B2 Retreat & Deliberately increase distance & Obtain an overview of cloud form or module relationships \\
B3 Enter / pass through & Enter or pass through spaces between modules & Observe the representation from another spatial position \\
B4 Circle & Move around a module to its other side & View flow, depth, or occlusion from another angle \\
B5 Change viewpoint & Lean, turn, look up or down, or noticeably reposition & Reveal hidden layers or display elements \\
B6 Touch & Contact permitted parts of the enclosure & Explore physical properties or boundaries \\
B7 Dwell / sustain viewing & Remain oriented toward a state & Continue inspecting the visible state \\
B8 Compare modules & Repeatedly view two or more modules & Compare locations or artistic weather states \\
B9 Wait for change & Observe a formation, movement, or dissipation event & Observe material formation, movement, or dissipation \\
B10 Point / read values & Point to or approach text, legends, or numbers & Inspect material appearance alongside digital explanation \\
B11 Discuss together & Talk, point, or compare with another viewer & Discuss observations with another viewer \\
\bottomrule
\end{tabular}
\end{table*}

\subsubsection{Self-Report Questionnaire}\label{sec:6.4.2}

Participants rated ten dimensions on study-specific 1-10 scales. We adapted mental demand, physical demand, temporal demand, effort, performance, and frustration from NASA-TLX \cite{hart1988tlx}. Spatial presence drew on the Presence Questionnaire \cite{witmer1998presence}, emotional arousal on the Self-Assessment Manikin \cite{bradley1994emotion}, and environmental action intention on the theory of planned behavior \cite{ajzen1991behavior}. Five project-specific items assessed perceived data comprehensibility. They measure participants' perceived understanding, not objectively tested comprehension.

\subsubsection{Open-Ended Responses}\label{sec:6.4.3}

After both conditions, participants answered eight open-ended questions. These covered what they found easiest or most difficult to understand, what appealed to or confused them, the roles of mist, light, cloud form, and spatial layout, environmental understanding, and their representational preferences. We transcribed the responses and checked them against the original forms. One researcher carried out two rounds of structured content analysis \cite{hsieh2005content}, first assigning descriptive codes close to participants' wording and then grouping them into themes. With only one coder, these themes provide context for participants' experiences but do not constitute a validated qualitative model.

\subsection{Data Analysis}\label{sec:6.5}

We paired questionnaire responses by participant ID and transformed reverse-scored items using \(11-x\) before calculating dimension scores. We compared conditions with two-sided paired \(t\)-tests and Holm correction across ten dimensions \cite{holm1979multiple}, using Wilcoxon signed-rank tests as robustness checks. We report paired Cohen's \(d_z\) and participant-level bootstrap 95\% confidence intervals, following the distinction between effect sizes for paired and independent designs \cite{lakens2013effect}. We also explored spatial behavior in the 13 participants' test recordings, totaling approximately 129.2 minutes. Each recording used a fixed camera view. ChatGPT assisted with analysis and plotting code. A pretrained YOLO11n model detected people in frames sampled at 1 Hz and resized to 640 \(\times\) 360 pixels. We selected candidates by bounding-box position, size, confidence, and temporal continuity. Low-confidence detections, ambiguous multi-person frames, and detections outside the position and size thresholds were marked invalid.

We estimated relative positions from each bounding box's horizontal center, height, and bottom edge, then mapped them to a shared schematic layout. We applied limited interpolation followed by five-point median and mean filters. These estimates cannot measure physical distance or speed because they are not calibrated to ground coordinates.

Figure~\ref{fig:13}(b) shows relative occupancy with equal participant weighting: each participant's valid samples contribute a total weight of one. We summed their two-dimensional position histograms and applied Gaussian smoothing, producing a heatmap of relative sample concentration rather than cumulative dwell time. Figure~\ref{fig:13}(c) connects estimates in temporal order for all 13 participants and overlays archived representative event markers. Figure~\ref{fig:13}(d) shows transitions among the start area, cloud modules, and side display. We merged brief label changes and collapsed consecutive repeated labels into region-visit sequences. Each change between adjacent regions counts as a directed transition, with pooled counts shown by edge weights and arrow widths. The network describes estimated changes of region; it does not measure attention, comparison, or comprehension.

\section{Results}\label{sec:7}

\subsection{Questionnaire Results}\label{sec:7.1}

\begin{figure*}[!t]
\centering
\includegraphics[width=\linewidth,keepaspectratio]{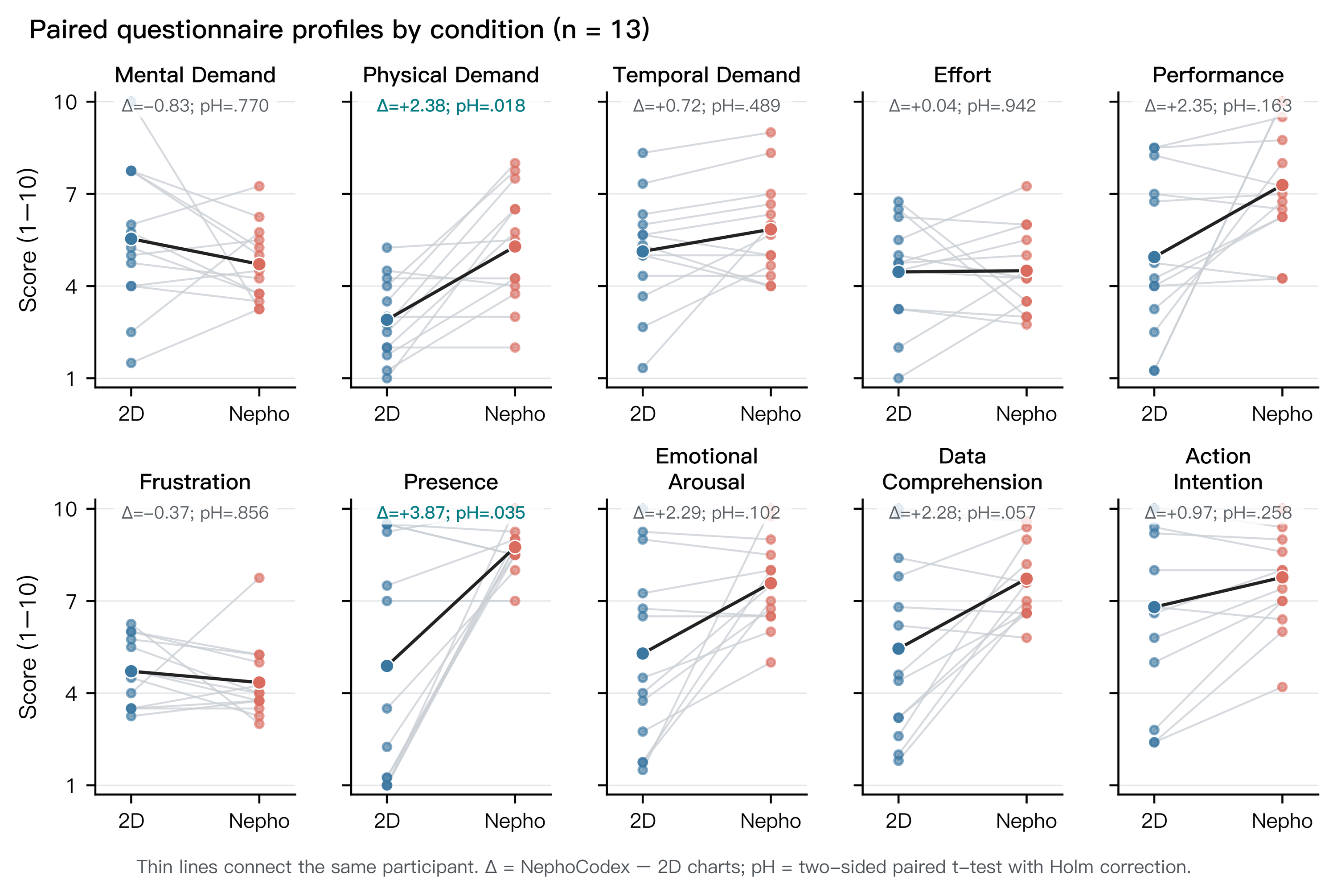}
\caption{Individual paired questionnaire ratings for Weather Charts and NephoCodex (n = 13). Points and connecting lines show participant-level variation.}
\label{fig:12}
\Description{Paired questionnaire ratings for thirteen participants across ten measures.}
\end{figure*}

NephoCodex produced the clearest differences in physical demand and spatial presence (Figure~\ref{fig:12}; Table~\ref{tab:5}). Physical demand increased from 2.90 to 5.29 (\(\Delta\) = 2.38, 95\% CI [1.08, 3.69], \(d_z\) = 1.10, Holm-adjusted p = .018); spatial presence increased from 4.88 to 8.75 (\(\Delta\) = 3.87, 95\% CI [1.50, 6.23], \(d_z\) = .99, Holm-adjusted p = .035). Perceived data comprehensibility also increased by 2.28 points, but did not reach the Holm-adjusted significance threshold (p = .057).

No other dimension showed a reliable difference after correction. Wilcoxon tests were more conservative, with Holm-adjusted p-values of .059 for physical demand and .105 for spatial presence. Increased spatial presence and physical demand were therefore supported under the prespecified paired t-test analysis. Differences in perceived comprehensibility and the other dimensions remain suggestive rather than conclusive.

\begin{table*}[!t]
\caption{Within-participant questionnaire comparisons (n = 13; 1-10 scales). Confidence intervals for \(\Delta\) are unadjusted; p-values are Holm-adjusted across ten paired t-tests.}
\label{tab:5}
\centering
\footnotesize
\setlength{\tabcolsep}{3pt}
\renewcommand{\arraystretch}{1.22}
\begin{tabular}{@{}>{\raggedright\arraybackslash}p{\dimexpr 0.15000\linewidth-1.667\tabcolsep\relax}>{\raggedright\arraybackslash}p{\dimexpr 0.14000\linewidth-1.667\tabcolsep\relax}>{\raggedright\arraybackslash}p{\dimexpr 0.16000\linewidth-1.667\tabcolsep\relax}>{\raggedright\arraybackslash}p{\dimexpr 0.20000\linewidth-1.667\tabcolsep\relax}>{\raggedright\arraybackslash}p{\dimexpr 0.23000\linewidth-1.667\tabcolsep\relax}>{\raggedright\arraybackslash}p{\dimexpr 0.12000\linewidth-1.667\tabcolsep\relax}@{}}
\toprule
\textbf{Dimension} & \textbf{Weather Charts M (SD)} & \textbf{NephoCodex M (SD)} & \textbf{\(\Delta\) [95\% CI]} & \textbf{\(d_z\) [bootstrap 95\% CI]} & \textbf{Holm p} \\
\midrule
Mental demand & 5.54 (2.35) & 4.71 (1.25) & \(-\).83 [\(-\)2.34, .69] & \(-\).33 [\(-\).92, .22] & .770 \\
Physical demand & 2.90 (1.32) & 5.29 (1.91) & 2.38 [1.08, 3.69] & 1.10 [.64, 2.03] & .018 \\
Temporal demand & 5.13 (1.85) & 5.85 (1.59) & .72 [\(-\).22, 1.66] & .46 [\(-\).03, 1.01] & .489 \\
Effort & 4.46 (1.71) & 4.50 (1.38) & .04 [\(-\)1.09, 1.17] & .02 [\(-\).51, .76] & .942 \\
Performance & 4.94 (2.62) & 7.29 (1.91) & 2.35 [.31, 4.38] & .70 [.36, 1.16] & .163 \\
Frustration & 4.71 (1.10) & 4.35 (1.25) & \(-\).37 [\(-\)1.34, .61] & \(-\).23 [\(-\)1.15, .28] & .856 \\
Spatial presence & 4.88 (3.82) & 8.75 (.78) & 3.87 [1.50, 6.23] & .99 [.53, 1.84] & .035 \\
Emotional arousal & 5.29 (3.02) & 7.58 (1.48) & 2.29 [.54, 4.04] & .79 [.45, 1.33] & .102 \\
Data comprehensibility & 5.45 (2.91) & 7.72 (1.31) & 2.28 [.74, 3.81] & .90 [.46, 1.66] & .057 \\
Action intention & 6.80 (2.94) & 7.77 (1.68) & .97 [\(-\).01, 1.95] & .60 [.16, 1.13] & .258 \\
\bottomrule
\end{tabular}
\end{table*}

\subsection{Spatial and Embodied Interaction}\label{sec:7.2}

\begin{figure*}[!t]
\centering
\includegraphics[width=\linewidth,keepaspectratio]{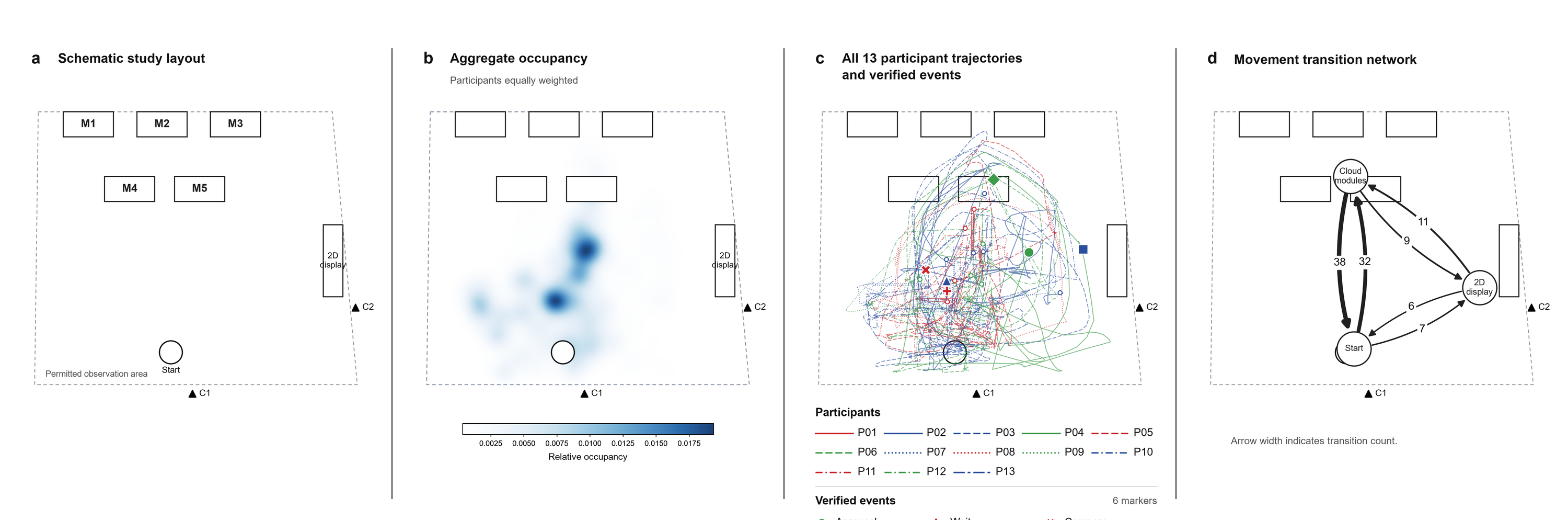}
\caption{Overview of participants' spatial behavior: (a) study layout, (b) aggregate occupancy, (c) trajectories and verified events for 13 participants, and (d) movement transition network. This figure illustrates spatial behavior rather than directly measuring comprehension.}
\label{fig:13}
\Description{Study layout, aggregate occupancy, participant trajectories and a movement-transition network.}
\end{figure*}

Participants generally stayed in one viewing position while reading values, colors, and legends in the Weather Charts condition. With NephoCodex, they approached transparent displays, changed viewpoint, moved among modules, compared states, and waited for clouds to form or dissipate. Figure~\ref{fig:13} summarizes their spatial distributions and trajectories.

\begin{figure*}[!t]
\centering
\includegraphics[width=\linewidth,keepaspectratio]{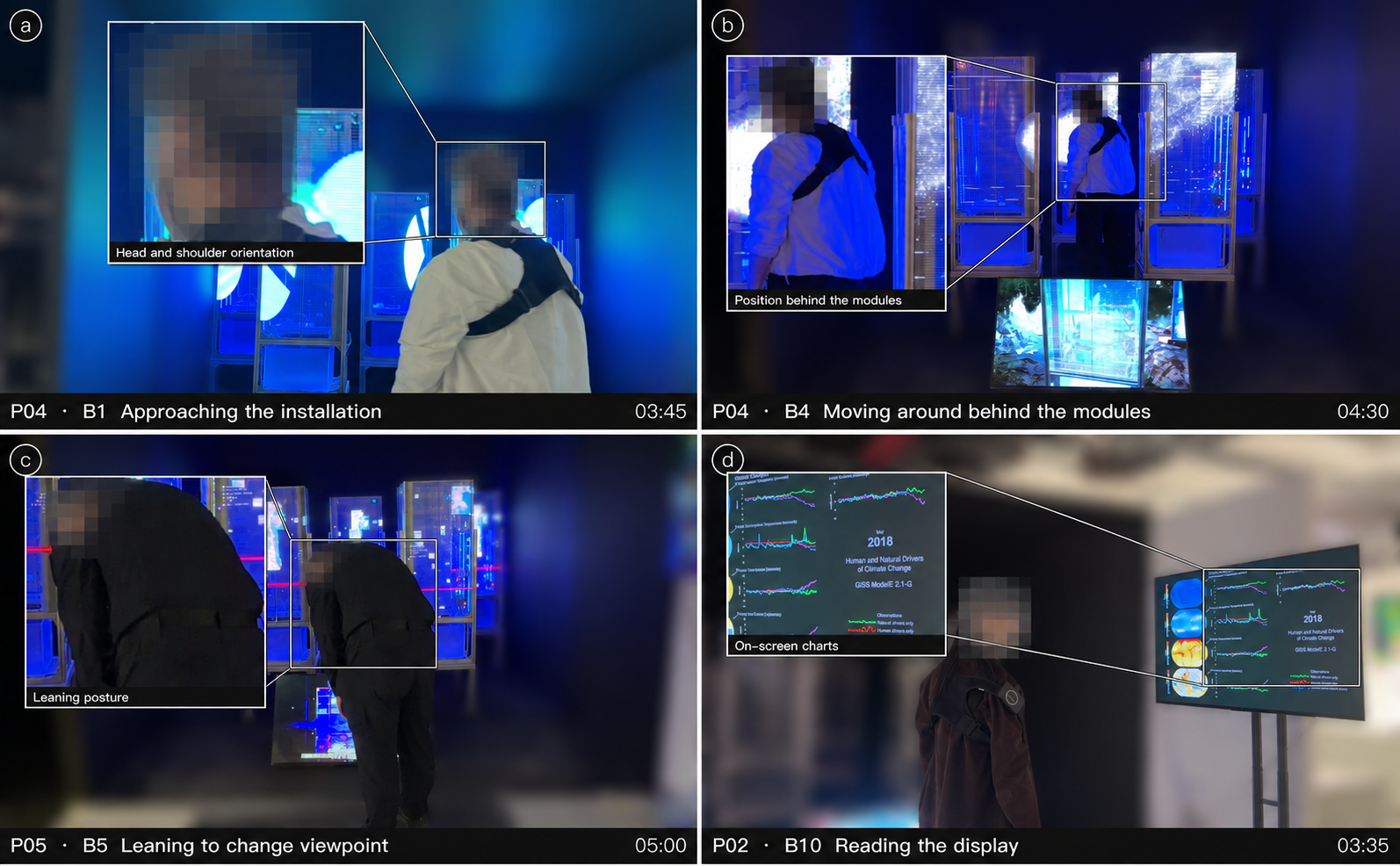}
\caption{Representative observable behaviors with detail insets: (a) P04 approaches the installation, (b) P04 circles behind a module, (c) P05 leans to change viewpoint, and (d) P02 reads a display. Frames retain original timestamps; insets show orientation, posture, and reading targets.}
\label{fig:14}
\Description{Four video frames show approaching, circling, leaning and reading the display, with detail insets.}
\end{figure*}

\begin{figure}[!htbp]
\centering
\includegraphics[width=\linewidth,keepaspectratio]{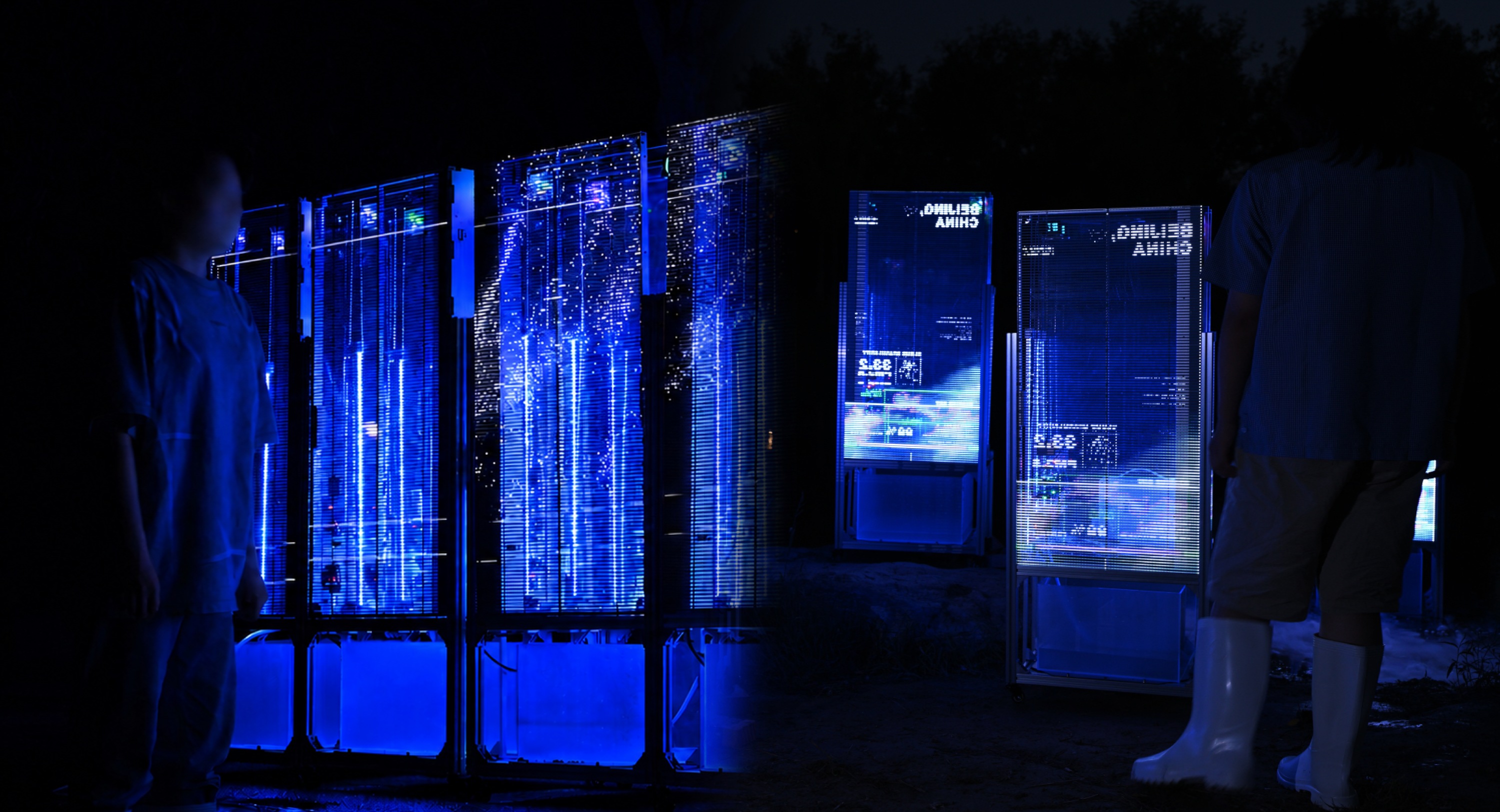}
\caption{Viewing and interaction during the NephoCodex exhibition, illustrating spatial relationships between viewers and transparent modules.}
\label{fig:15}
\Description{Exhibition viewers interacting with the transparent NephoCodex modules.}
\end{figure}

We observed four recurring interaction patterns with NephoCodex. Approach-and-read events involved moving from watching mist to inspecting transparent displays, linking changes to values. In move-and-reveal events, participants moved laterally to expose elements hidden from their initial viewpoint. Wait-and-compare events involved watching a cloud form or dissipate before comparing it with another module. During occlusion-repair events, participants repositioned because text, mist, or spatial depth impeded reading. Figure~\ref{fig:14} shows representative behaviors and details; Figure~\ref{fig:15} shows viewing during the exhibition.

These spatial and temporal viewing behaviors were largely absent from the chart condition. Some allowed participants to observe material transitions, while others helped them overcome occlusion or poor legibility.

\begin{table*}[!t]
\caption{Observable actions and their interaction roles.}
\label{tab:6}
\centering
\small
\setlength{\tabcolsep}{3pt}
\renewcommand{\arraystretch}{1.22}
\begin{tabular}{@{}>{\raggedright\arraybackslash}p{.18\linewidth}>{\raggedright\arraybackslash}p{.29\linewidth}>{\raggedright\arraybackslash}p{.13\linewidth}>{\raggedright\arraybackslash}p{\dimexpr.40\linewidth-6\tabcolsep\relax}@{}}
\toprule
\textbf{Action} & \textbf{Observable criterion} & \textbf{Target} & \textbf{Observed interaction role} \\
\midrule
Approach & Reduce distance & Display & Read precise values \\
Change viewpoint & Reposition laterally & Mist / display & View depth or reveal hidden layers \\
Wait & Remain oriented during a transition & Mist & Observe formation / dissipation \\
Circle & Move to the other side & Module & Inspect airflow / depth \\
Repeatedly reposition & Search for a legible label & Display & Resolve occlusion \\
\bottomrule
\end{tabular}
\end{table*}

\subsection{Open-Ended Responses}\label{sec:7.3}

The 104 open-ended responses yielded four recurring themes (Table~\ref{tab:7}).

T1: Weather charts helped participants read trends and make comparisons, but required them to decode symbols. Participants valued the compact overview while finding some legends, line symbols, and relationships among graphical elements difficult to interpret.

T2: NephoCodex emphasized process and spatial presence. Participants described changing cloud states, immersion, and multisensory experience, and valued what mist, light, or spatial elements contributed. Several characterized these benefits as enjoyment or atmosphere rather than understanding.

T3: Legibility and source clarity remained difficult. Participants had trouble reading transparent displays, following rapid changes, identifying imagery sources, extracting precise values, and comparing modules at different depths. P11 described it as “difficult to capture an intuitive data value,” while P7 was unsure about “the source or reason for the imagery presented by the installation.”

T4: Preferences depended on the representation's role. Participants often favored NephoCodex for aesthetics, novelty, enjoyment, or immersion rather than analytical accuracy. P10 preferred conventional visualization for everyday use and NephoCodex for exhibitions. P7 found the physical experience direct and immersive, while using charts to confirm and supplement information.

Evidence of environmental awareness was limited. Nine participants gave affirmative or partly affirmative responses, but only two described specific relational insights. P3 discussed the interconnectedness of global climate change, and P7 described relationships between human and natural influences on weather. Three responses primarily expressed concern, urgency, or emotional engagement; four participants could not specify what had changed; three reported no change; and one response was indeterminate. Some participants thus articulated environmental insights, but their accounts do not establish a general change in understanding or attitudes. This limit accords with calls to assess environmental outcomes beyond usability or favorable experience \cite{remy2018evaluation}.

\begin{table*}[!t]
\caption{Themes from structured content analysis of open-ended responses. Counts indicate participants associated with each response pattern; themes are not mutually exclusive.}
\label{tab:7}
\centering
\small
\setlength{\tabcolsep}{3pt}
\renewcommand{\arraystretch}{1.22}
\begin{tabular}{@{}>{\raggedright\arraybackslash}p{\dimexpr 0.24000\linewidth-1.500\tabcolsep\relax}>{\raggedright\arraybackslash}p{\dimexpr 0.20000\linewidth-1.500\tabcolsep\relax}>{\raggedright\arraybackslash}p{\dimexpr 0.28000\linewidth-1.500\tabcolsep\relax}>{\raggedright\arraybackslash}p{\dimexpr 0.28000\linewidth-1.500\tabcolsep\relax}@{}}
\toprule
\textbf{Theme} & \textbf{Coverage} & \textbf{Supporting observations} & \textbf{Countervailing considerations} \\
\midrule
T1 Charts support trends and comparisons & 11/13 mentioned trends; 9/13 mentioned symbolic burden & Temporal curves support overview and comparison. & Legends and dense lines require decoding. \\
T2 Physical clouds foreground process and presence & 7/13 mentioned change; 4/13 mentioned immersion or multisensory cues & Mist and light make change spatial and engaging. & Enjoyment or immersion does not establish correct understanding. \\
T3 Legibility and source clarity impose costs & 10/13 reported at least one difficulty & Values, sources, timing, and spatial comparisons may be unclear. & 3/13 explicitly reported no difficulty. \\
T4 Preferences reveal complementary roles & 9 preferred the installation; 2 preferred charts; 2 expressed mixed or contextual preferences & The installation is appealing and experiential. & Charts remain preferable for everyday, precise, or comparative reading. \\
\bottomrule
\end{tabular}
\end{table*}

\section{Discussion}\label{sec:8}

Turning weather data into physical clouds introduces variation at several points. Predictions are probabilistic, mist does not reproduce exactly identical states, and viewers may find an appealing display ambiguous. NephoCodex shows how these forms of uncertainty differ as data pass through computation, material realization, and interpretation.

\subsection{From Computational Uncertainty to Material Variability}\label{sec:8.1}

NephoCodex maps multivariate weather time series to probability distributions over artistic states. At this computational stage, the distributions retain ambiguity among candidate states and give it numerical form that can be represented, compared, and logged.

The resulting material behavior varies in ways that a probability distribution cannot specify. Before execution, the controller translates predicted states into commands that entropy, environmental sensing, and safety rules may modify. After execution, airflow, droplets, condensation, and visitor disturbances continue to shape the mist. The same computational state can therefore produce different physical outcomes.

Computational uncertainty and material variability can be controlled in different ways. Probabilities, entropy, and logged adjustments make computational uncertainty explicit. Material behavior cannot be reduced to fully repeatable visual states, and our design does not require exact repetition. Instead, designers can constrain the conditions under which materials respond. The system can explain why it issued a command even when it cannot reproduce each subsequent eddy or droplet configuration.

This distinction extends accounts of materiality in tangible and embodied interaction \cite{dourish2001action,hornecker2006tangible}. Material properties affect how viewers encounter and interpret a physicalization. Mist occupies space, changes over time, and responds to its surroundings, making these effects visible. The material between a computational representation and its perceptible form can therefore behave in ways the data alone do not determine.

\subsection{Experiential Consequences of Material Physicalization}\label{sec:8.2}

The clearest quantitative changes were increased spatial presence and physical demand. Mist's depth, occlusion, direction, and duration also elicited behaviors largely absent from Weather Charts. Participants approached transparent displays, changed viewpoint, moved among modules, compared states, and waited for formation or dissipation. These physical and temporal properties affected where participants stood and how they attended to the representation.

Interpreting physical demand requires examining why participants moved. Sometimes they inspected depth or airflow, waited to observe formation and dissipation, or crossed between modules to compare states. At other times, they repeatedly repositioned to find a legible value or see past an occlusion. Physical demand reflected both engagement with the representation and effort spent compensating for interface limitations.

Increased spatial presence and physical demand did not establish improved understanding. The difference in perceived comprehensibility was suggestive but sensitive to correction. Self-reported effort, mental demand, and frustration showed no reliable condition differences. Participants felt more situated in the representation and interacted with it more physically, but we did not demonstrate an increase in understanding.

Participants' responses help explain the difference between experience and understanding. Compact weather charts supported numerical reading, trend identification, and comparison. NephoCodex emphasized process, spatial variation, and multisensory experience. Yet participants struggled to extract precise values, trace the sources of changes, follow rapid transitions, and compare modules at different depths. Ten of the thirteen participants reported at least one of these difficulties.

These findings support pairing experiential and analytical representations. Material behavior can convey intensity, transitions, duration, and spatial change. A stable digital layer can identify variables, values, times, sources, mappings, and other information that mist may obscure. P7 described the physical experience as direct and immersive and valued conventional charts for “confirming and supplementing information.” The two displays thus served complementary roles in this participant's account.

NephoCodex may shift attention from symbolic decoding toward movement, sensory integration, and interpreting changing materials. Subjective workload, preference, and behavioral involvement alone cannot establish that shift. Our evidence shows a change in how participants physically encounter the representation; it does not demonstrate a general improvement in cognitive engagement or understanding.

\subsection{Bounded Material Agency}\label{sec:8.3}

Bounded material agency describes how computational control coexists with material variability. In NephoCodex, computation constrains the conditions in which mist forms while leaving its exact visible behavior variable. The mist remains subject to control, but its appearance is only partly determined by that control.

Machine learning maps multivariate weather time series to distributions over artistic states. These distributions make predictive confidence and class imbalance visible where deterministic assignments would conceal them. They propose controls that remain subject to further regulation, leaving the final cloud unspecified.

Entropy, environmental sensing, and safety rules can alter commands before execution. Airflow, droplets, condensation, and visitor disturbances continue to shape the mist afterward. The cloud's appearance depends on this combination of designed regulation and physical variation; neither the model nor autonomous material agency alone explains it.

The system controls actuator parameters, responds to environmental conditions, enforces safety constraints, and logs decisions. It cannot determine every eddy, droplet configuration, or transient cloud boundary. Bounded material agency places control in the conditions that produce material behavior, while allowing the resulting appearance to vary.

Probabilities, entropy, and logged adjustments describe computational uncertainty. Material variability persists after these decisions, so keeping the two distinct allows us to explain a command without claiming to reproduce every physical configuration. Accounting for the computation identifies which variations arise from designed regulation and which emerge as the material responds.

Machine learning is useful here because the mapping involves multivariate temporal structure and probabilistic mixtures of material states. It is not a requirement for data physicalization or bounded material agency: a transparent deterministic mapping may be simpler and easier to interpret when it suffices. Machine learning contributes when probabilities matter to the representation and its outputs meaningfully constrain material expression without directly controlling the final form.

\subsection{Design Implications for Variable Material Representations}\label{sec:8.4}

First, distinguish computational uncertainty from material variability. NephoCodex shows that probabilistic model outputs, uncertainty-based adjustments, and variable physical outcomes arise through different processes. Keeping these processes distinct clarifies what computation determines and what remains contingent as the material responds.

Second, keep the relationship between computational states and actuator behavior consistent while allowing visible outcomes to vary. Exact reproduction of a dynamic medium such as mist may be infeasible or unnecessary. Actuator ranges, environmental sensing, and safety mechanisms can constrain the conditions in which materials respond while allowing variation within those limits.

Third, examine why a physicalization demands bodily effort. Movement, viewpoint changes, and waiting may reveal spatial or temporal properties, or compensate for occlusion, poor legibility, and difficult comparisons. Designers should distinguish actions that reveal information about the material from effort imposed by avoidable interface constraints.

Fourth, keep stable analytical information alongside variable materials. States, values, times, sources, mappings, and explanations help viewers read information they cannot recover from mist alone. Material behavior can convey process and change; digital information can support numerical reading, provenance, and comparison. This combination is especially useful when transient media occlude information or require several viewpoints to inspect.

Fifth, make computational regulation inspectable even when material outcomes cannot be reproduced exactly. Logs of probabilistic predictions, entropy-based adjustments, environmental attenuation, and safety interventions let designers trace how the system arrived at an actuator command. This trace does not give every feature of the resulting mist a single computational cause.

Finally, evaluate experience and comprehension separately. Presence, preference, movement, and multisensory engagement describe how physicalization changes an experience. They do not establish better understanding, and perceived comprehensibility also differs from objectively tested comprehension. Evaluations should measure these outcomes separately so that embodiment or engagement does not stand in for interpretive accuracy.

A physical representation can retain temporal and material variation while remaining designed and regulated. Explicit computational constraints and analytical support make this possible. Designers need to specify what must remain controlled, what may vary, and how to communicate and evaluate the difference.

\section{Limitations and Future Work}\label{sec:9}

The small formative study provides situated design evidence rather than population estimates, and its disciplinary breadth does not establish saturation. The user study involved only 13 participants at one outdoor site, many with art or interaction-design backgrounds. It was not preregistered, and the study-specific self-report dimensions lack independent psychometric validation. Objective weather-comprehension tasks are needed to assess accuracy; perceived comprehensibility cannot establish it. Behavioral coding identified movement, viewpoint change, waiting, and comparison. Future work should quantify their frequency, duration, and transitions across conditions and test how they relate to interpretation and comprehension.

We evaluated the model at one location over one historical interval, with limited recall for the rare convective-alert state. The implementation also lacks a synchronized evaluation of end-to-end latency and the correspondence between commands and visible mist. Technical evaluation should test performance across locations and seasons, calibrated actuator responses, and synchronized closed-loop behavior. Sustained public deployment requires assessment of resource consumption, maintenance, and accessibility. Studies should also directly test whether viewers perceive or interpret computational uncertainty and material variability, since probabilistic control alone cannot establish this. The physicalization's life cycle warrants assessment of water and energy use, maintenance, reuse, and disposal \cite{morais2024sustainability}. We did not quantify these costs.

\section{Conclusion}\label{sec:10}

NephoCodex translates predicted weather-state distributions into proposed controls for mist, airflow, and light. Entropy-based regulation, local sensing, and safety constraints govern execution while leaving the mist's exact form variable. The system demonstrates how computational uncertainty can influence material variability in a dynamic physical representation.

In a within-participant study, NephoCodex increased spatial presence and physical demand. Participants moved, changed viewpoint, and waited to observe the representation. Perceived comprehensibility remained inconclusive after multiple-comparison correction, and self-reported effort and mental demand showed no reliable differences between conditions. Greater spatial presence did not establish greater understanding or cognitive engagement.

Bounded material agency describes this relationship between computational constraint and variable material behavior. Uncertainty can guide control while leaving appearance partly undetermined. Variable materials can make process and change perceptible, but also introduce ambiguity. NephoCodex supports pairing material encounters with digital information about values, sources, mappings, and comparisons. In this approach, uncertainty helps set the conditions in which materials respond and remains part of the physicalization.

\FloatBarrier
\bibliographystyle{ACM-Reference-Format}
\bibliography{reference}

\clearpage
\onecolumn
\appendix
\renewcommand{\thetable}{A\arabic{table}}
\setcounter{table}{0}
\section{Features and Cluster Centers for State Construction}\label{sec:A}

\begin{table}[!htbp]
\caption{Descriptors of the next three hours used for clustering. These descriptors construct target labels and are not inputs at prediction time.}
\label{tab:A1}
\centering
\small
\setlength{\tabcolsep}{3pt}
\renewcommand{\arraystretch}{1.22}
\begin{tabular}{@{}>{\raggedright\arraybackslash}p{\dimexpr 0.30000\linewidth-1.000\tabcolsep\relax}>{\raggedright\arraybackslash}p{\dimexpr 0.70000\linewidth-1.000\tabcolsep\relax}@{}}
\toprule
\textbf{Feature} & \textbf{Definition} \\
\midrule
Future cloud mean & Mean cloud cover over the next 3 h \\
Future cloud maximum & Maximum cloud cover over the next 3 h \\
Future precipitation total & Accumulated precipitation over the next 3 h \\
Future precipitation maximum & Maximum hourly precipitation over the next 3 h \\
Future humidity mean & Mean relative humidity over the next 3 h \\
Future wind maximum & Maximum wind speed over the next 3 h \\
Future gust maximum & Maximum gust speed over the next 3 h \\
Future pressure change & Pressure at +3 h minus pressure at the current hour \\
\bottomrule
\end{tabular}
\end{table}

\begin{table}[!htbp]
\caption{Cluster centers for the five weather states. Cluster identifiers follow model outputs; units are given in column headings.}
\label{tab:A2}
\centering
\footnotesize
\setlength{\tabcolsep}{3pt}
\renewcommand{\arraystretch}{1.22}
\begin{tabular}{@{}>{\raggedright\arraybackslash}p{\dimexpr 0.07000\linewidth-1.800\tabcolsep\relax}>{\raggedright\arraybackslash}p{\dimexpr 0.16000\linewidth-1.800\tabcolsep\relax}>{\raggedright\arraybackslash}p{\dimexpr 0.10000\linewidth-1.800\tabcolsep\relax}>{\raggedright\arraybackslash}p{\dimexpr 0.10000\linewidth-1.800\tabcolsep\relax}>{\raggedright\arraybackslash}p{\dimexpr 0.10000\linewidth-1.800\tabcolsep\relax}>{\raggedright\arraybackslash}p{\dimexpr 0.10000\linewidth-1.800\tabcolsep\relax}>{\raggedright\arraybackslash}p{\dimexpr 0.10000\linewidth-1.800\tabcolsep\relax}>{\raggedright\arraybackslash}p{\dimexpr 0.09000\linewidth-1.800\tabcolsep\relax}>{\raggedright\arraybackslash}p{\dimexpr 0.09000\linewidth-1.800\tabcolsep\relax}>{\raggedright\arraybackslash}p{\dimexpr 0.09000\linewidth-1.800\tabcolsep\relax}@{}}
\toprule
\textbf{Cluster} & \textbf{Regime} & \textbf{Cloud mean \%} & \textbf{Cloud max \%} & \textbf{Precip sum mm} & \textbf{Precip max mm} & \textbf{Humidity mean \%} & \textbf{Wind max} & \textbf{Gust max} & \textbf{Pressure \(\Delta\) hPa} \\
\midrule
0 & calm/open & 37.40 & 48.18 & 0.23 & 0.19 & 83.43 & 2.62 & 5.67 & +0.27 \\
1 & rain-bearing & 91.92 & 96.36 & 0.54 & 0.34 & 89.33 & 1.87 & 3.56 & +0.43 \\
2 & convective alert & 94.04 & 97.18 & 8.92 & 5.55 & 82.61 & 2.61 & 6.11 & \(-\)0.97 \\
3 & fragmented transition & 88.59 & 94.11 & 0.94 & 0.64 & 74.13 & 3.40 & 7.41 & \(-\)1.67 \\
4 & layered/overcast & 93.35 & 97.16 & 0.55 & 0.31 & 86.25 & 3.46 & 6.43 & +0.77 \\
\bottomrule
\end{tabular}
\end{table}
\end{document}